\documentclass[aps,twocolumn,showpacs,preprintnumbers,amsmath,amssymb,nofootinbib,superscriptaddress,showkeys]{revtex4}

\usepackage{epsfig}
\usepackage{graphicx}

\def\u{{\bf u}} \def\U{{\bf U}} \def\S{{\bf S}} \def\h{{\bf h}}
\def\L{{\bf L}} \def\E{{\bf 1}} \def\j{{\bf j}} \def\y{{\bf y}}
\def\M{{\bf M}} 
\def\f{{\bf f}}

\begin{document}

\title{Wigner symmetry, Large $N_c$ and Renormalized One Boson Exchange
Potentials.}

\author{A. Calle Cord\'on}\email{alvarocalle@ugr.es}
  \affiliation{Departamento de F\'isica At\'omica, Molecular y
  Nuclear, Universidad de Granada, E-18071 Granada, Spain.}
  \affiliation{Department of Physics, U-3046, University of
  Connecticut, Storrs, CT, 06269-3046.}
\author{E. Ruiz
  Arriola}\email{earriola@ugr.es}
  \affiliation{Departamento de F\'isica At\'omica, Molecular y Nuclear,
  Universidad de Granada,
  E-18071 Granada, Spain.}
\date{\today}

\begin{abstract} 
\rule{0ex}{3ex} Wigner symmetry in Nuclear Physics provides a unique
example of a non-perturbative long distance symmetry, a symmetry
strongly broken at short distances. We analyse the consequences of
such a concept within the framework of One Boson Exchange Potentials
in NN scattering and keeping the leading $N_c$
contributions. Phenomenologically successful relations between singlet
$^1S_0$ and triplet $^3S_1$ scattering phase shifts are provided in
the entire elastic region. We establish symmetry breaking relations
among non-central phase shifts which are successfully fulfilled by
even-L partial waves and strongly violated by odd-L partial waves, in
full agreement with large $N_c$ requirements.  
\end{abstract}

\maketitle



\section{Introduction}

Symmetries have traditionally been very useful in nuclear physics
partly because the force at the hadronic level is not well known at
short distances~\cite{1941RPPh....8..274W,Wilkinson:1969,
1999RPPh...62.1661V}.  In some cases, like isospin, chiral or heavy
quark symmetry, the invariance can directly be traced from the
fundamental QCD Lagrangean and formulated in terms of the underlying
quark and gluonic degrees of freedom. In some other cases the
connection is less straightforward. Many years ago Wigner and Hund
proposed~\cite{Wigner:1936dx,1937ZPhy..105..202H} extending the spin
and isospin $SU_S (2) \otimes SU_I (2) $ symmetry into the larger
$SU(4)$ group where the nucleon-spin states $p\uparrow$,
$p\downarrow$, $n\uparrow$, $n\downarrow$ correspond to the
fundamental representation, and hence providing a supermultiplet
structure of nuclear energy levels as well as new selection rules for
nuclear transitions and response
functions~\cite{1970AnPhy..60..209D}. The corresponding $SU(4)$ mass
formula was found to be at least as good as the well known
Weizs\"acker one ~\cite{Cauvin:1980sj,VanIsacker:1997vh}. Spin-orbit
interaction of the shell model obviously violate the symmetry, and
indeed a breakdown of $SU(4)$ has been reported for heavier
nuclei~\cite{Vogel:1993zz} while nuclear matter has been addressed
in~\cite{Nayak:2001tg}. Double binding energy differences have been
shown to be a useful test of the symmetry~\cite{VanIsacker:1995ux}.
Recently, inequalities for light nuclei based on $SU(4)$ and Euclidean
path integrals have been derived by neglecting all but S-wave
interactions~\cite{Chen:2004rq}.

Despite its relative success along the years, $SU(4)$ symmetry has
been treated as an accidental one within the traditional approach to
Nuclear Physics and guessing its origin from QCD has been a subject of
some interest in the last decade. Indeed, attempts to justify $SU(4)$
spin-flavour symmetry from a more fundamental level have been carried
out along several lines. Based on the limit of large number of colors
$N_c$ of QCD~\cite{'tHooft:1973jz,Witten:1979kh}, it was
shown~\cite{Kaplan:1995yg,Kaplan:1996rk} that if the nucleon momentum
scales as $p \sim N_c^0$, the nuclear potentials scale either as $N_c$
or $1/N_c$, depending upon the particular spin-isospin channel, which
shows that the NN force could be determined with $1/N_c^2$ relative
accuracy. It was found that the leading potential would be $SU(4)$
symmetric if the tensor force was neglected in addition, a plausible
assumption for light nuclei where S-waves dominate. Although these
estimates are conducted directly in terms of quarks and gluons,
quark-hadron duality allows to reformulate these results in terms of
purely hadronic degrees of freedom, providing a rationale for the
One-Boson-Exchange (OBE) potential models~\cite{Machleidt:1987hj}, and
the internal consistency of Two-~\cite{Banerjee:2001js} and Multiple
Boson Exchanges~\cite{Belitsky:2002ni,Cohen:2002im}. The analysis of
sizes of volume integrals of phenomenologically successful potentials
confirms the large $N_c$ expectations~\cite{Riska:2002vn}. The large
size of scattering lengths was regarded as a fingerprint of the
$SU(4)$ symmetry within an Effective Field Theory (EFT)
viewpoint~\cite{Mehen:1999qs} using the Power Divergent Subtracted
(PDS) scheme; singlet and triplet renormalized couplings coincide at
the natural renormalization scale $ \mu \sim m_\pi \gg 1/ \alpha_s ,
1/ \alpha_t $, with $\alpha_s$ and $\alpha_t$ the scattering lengths,
and a contact interaction makes sense in such a scaling regime.
Resonance saturation based on the elimination of exchanged mesons in
the OBE Bonn potential~\cite{Machleidt:1987hj} at very low energies
was also shown to reproduce the EFT approach and to agree numerically
with the Wigner symmetry
expectations~\cite{Epelbaum:2001fm}. According to Refs.~\cite{
Epelbaum:2002gb,Beane:2002xf,Braaten:2003eu,Epelbaum:2006jc,Hammer:2007kq}
QCD might be close to a point where the effective theory had an
$SU(4)$ symmetry at zero energy as well as discrete scale invariance
if the pion mass was larger than its physical value, around $m_\pi
\sim 200 {\rm MeV}$.  This nice idea might be confirmed by recent
fully dynamical lattice QCD determinations of the scattering
lengths~\cite{Beane:2006mx} and quenched lattice QCD evaluations of NN
potentials~\cite{Ishii:2006ec,Aoki:2008hh} where indeed unphysical
pion masses are probed.

While the proclaimed symmetry holds in a range where scale invariance
sets in and EFT methods based on contact interactions can be
applied~\cite{Mehen:1999qs,Epelbaum:2001fm}, is not obvious {\it what}
are the implications for the lightest NN system itself for finite
energies and for physical pion masses.  In particular, the scale
dependence of the contact interaction is modified when the finite
range of the long distance potential is taken into account. To be
specific, low energy NN scattering is dominated by S-waves in
different $(S,T)$ channels where spin and isospin are interchanged,
$(1,0)$=$^1S_0 \leftrightarrow (0,1)$= $^3S_1$.  Wigner $SU(4)$
symmetry predicts {\it identical} interactions in both $^1S_0$ and
$^3S_1$ channels.  The above mentioned identity of the $^1S_0$ and
$^3S_1$ potentials holds also in the large
$N_c$-expansion~\cite{Kaplan:1995yg,Kaplan:1996rk}, so we take
advantage of this fact by using the leading $N_c$-OBE potentials which
simplifies the discussion to a large extent as we discuss in
Sect.~\ref{sec:OBE}. In contrast, the corresponding phase shifts from
Partial Wave Analyses~\cite{Stoks:1993tb} are {\it very different} at
all energies. We are thus confronted with an intriguing puzzle since
it is not obvious at all in what sense should the symmetry be
interpreted for the NN system; it would be difficult to understand
otherwise the successes of $SU(4)$ for light nuclei.  A second puzzle
arises from an embarrasing cohabitation of conflicts and agreements
between large-$N_c$ studies and Wigner symmetry.  Despite the initial
claim~\cite{Kaplan:1995yg} a more complete
analysis~\cite{Kaplan:1996rk} could only justify the Wigner symmetry
in even-L partial waves while for odd-L a violation of the symmetry
was expected. However, doing so required neglecting the tensor force,
which according to the Wigner symmetry should vanish, but it is a
leading contribution to the potential in the large $N_c$ limit. Thus,
while some pieces of the NN potential (such as e.g. spin-orbit) are
suppressed in both schemes, some others are not simultaneously small.
These conflicts between the time-honoured $SU(4)$ Wigner symmetry and
the QCD based large-$N_c$ expansion for odd-L channels require an
explanation and naturally pose the question on the validity of either
framework.

In the present work we analyze both puzzles by introducing the concept
of a {\it long distance symmetry} firstly to understand the meaning of
Wigner symmetry in those cases where its validity complies with large
$N_c$ expectations. This is a case where we expect the symmetry to be
more robust. Once this is done, it is pertinent to dilucidate the
validity of the symmetry in those cases where a possible conflict with
the large $N_c$ expansion arises. Our discussion is tightly linked to
the coordinate space renormalization discussed in previous
works~\cite{PavonValderrama:2005wv,PavonValderrama:2007nu}. This
approach while borrowing the physical motivation from EFT theories
provides a quantum mechanical framework where the emphasis is placed
on the non-perturbative aspects of the NN problem, a playground where
the standard EFT viewpoint has encountered notorious difficulties.
The method is reviewed in Sect.~\ref{sec:uni-renor} for completeness.
We find that for S-waves the Wigner symmetry holds in a much wider
range than the applicability of a contact interaction suggests if the
finite range of the interaction is incorpored. As a byproduct we
provide in Sect.~\ref{sec:numeric} {\it quantitative} predictions; the
seemingly independent triplet and singlet S-waves phase shifts
corresponding to iso-vector and iso-scalar states respectively for the
np system are shown to be neatly intertwined in the entire elastic
region. A similar correlation can also be established between the
$^1S_0$ virtual state and the $^3S_1$ deuteron bound state.  Actually,
we show how the symmetry may be visualized for large scattering
lengths due to the onset of scale invariance. Symmetry breaking due to
inclusion of further counter-terms, tensor interaction and spin-orbit
interaction are discussed in Sect.~\ref{sec:symm-break}. We show how a
sum rule for supermultiplet phase shifts splitting due to spin-orbit
and tensor interactions is well fulfilled for non-central L-even
waves, and strongly violated in L-odd waves where a Serber-like
symmetry holds instead. This pattern of $SU(4)$-symmetry breaking
complies to the large $N_c$ expectations, a somewhat unexpected
conclusion. Finally, in Sect.~\ref{sec:concl} we provide our main
conclusions and outlook for further work.

\section{OBE potentials, Large $N_c$ and Wigner symmetry}
\label{sec:OBE}

Our starting point is the field theoretical OBE model of the NN
interaction~\cite{Machleidt:1987hj} which includes all mesons with
masses below the nucleon mass, i.e., $\pi$, $\sigma(600), $$\eta$,
$\rho(770)$ and $\omega(782)$. For the purpose of discussing $SU(4)$
Wigner symmetry within the OBE framework (see Appendix~\ref{eq:su4}
for a brief overview) we will deal here with S-waves only, neglecting
for the moment the S-D wave mixing stemming from the tensor force as
required by Wigner symmetry. Our results of Sect.~\ref{sec:numeric}
and estimates in Sect.~\ref{sec:tensor} will provide the {\it a
posteriori} justification of this simplifying assumption. Non-central
waves and the role of spin-orbit as well as tensor force will be
treated in Sect.~\ref{sec:non-central} as $SU(4)$-breaking
perturbations. For the S-waves the NN potential reads
\begin{eqnarray}
V= V_C + \tau W_C  + \sigma V_S + \tau \sigma W_S  \, , 
\end{eqnarray} 
where $\tau= \tau_1 \cdot \tau_2 = 2 T(T+1)-3 $ and $\sigma= \sigma_1
\cdot \sigma_2 = 2 S (S+1)-3 $ and Pauli principle requires
$(-)^{S+T+L}=-1$.  Thus,  for the spin singlet $^1S_0$ and spin triplet
$^3S_1$ states we get
\begin{eqnarray}
V_s &=& V_C + W_C  - 3V_S  -3 W_S \, ,  \\
V_t &=& V_C -3 W_C  + V_S  -3 W_S \, , 
\end{eqnarray} 
To simplify the discussion we will discard terms in the potential
which are phenomenologically small. Actually, according to
Refs.~\cite{Kaplan:1995yg,Kaplan:1996rk} in the leading large $N_c$
one has $V_C \sim W_S \sim N_c$ while $V_S \sim W_C \sim 1/N_c$. In
terms of meson exchanges (see also Ref.~\cite{Banerjee:2001js}) one
has the contributions
\begin{eqnarray}
V_s (r) = V_t (r) &=& -\frac{ g_{\pi NN}^2 m_{\pi}^2} {16 \pi M_N^2}
\frac{e^{-m_{\pi} r}} {r} - \frac{ g_{\sigma NN} ^2}{4 \pi}\frac{e^{-
m_{\sigma} r}} {r}  
\nonumber \\ 
&+& \frac{g_{\omega NN}^2}{4 \pi}\frac{e^{-m_{\omega}
r}}{r} -
\frac{f_{\rho NN}^2 m_\rho^2}{8 \pi M_N^2 }\frac{e^{-m_{\rho}
r}}{r} \nonumber \\ 
&+&
{\cal O} \left( {N_c}^{-1} \right) \, , 
\label{eq:pot}
\end{eqnarray}
where $g_{\sigma NN}$ is a scalar type coupling, $g_{\pi NN}$ a
pseudo-scalar derivative coupling, $g_{\omega NN}$ is a vector
coupling and $f_{\rho NN}$ the tensor derivative coupling (see
\cite{Machleidt:1987hj} for notation).  Here, the scheme proposed
in~\cite{Partovi:1969wd} of neglecting both energy and nonlocal
corrections is realized explicitly. In principle the large $N_c$ limit
contains infinitely many multi-meson exchanges which decay
exponentially with the sum of the exchanged meson masses. However, NN
scattering in the elastic region below pion production threshold
probes CM momenta $p < p_{max} = 400$ MeV. Given the fact that
$1/m_{\omega} = 0.25 \mathrm{fm} \ll 1/p_{max} = 0.5\mathrm{fm}$ we
expect heavier meson scales to be irrelevant, an in particular
$\omega$ and $\rho$ themselves, are expected to be at most marginally
important~\footnote{This of course does not exclude explicit and
leading $N_c$ uncorrelated multiple pion exchanges at, i.e. background
non-resonant contributions in $\pi\pi$ or $\pi \rho$ scattering. We
expect them not to be dominant once $\sigma$, $\rho$ and $\omega$ are
included.}. Note that, in any case, when $m_\omega=m_\rho$ the
redundant combination $g_{\omega NN}^2 - f_{\rho NN}^2 m_\rho^2 / (2
M_N^2)$ appears, indicating a further source of cancellation between
$\rho$ and $\omega$ in this channel. Moreover, since the leading
contributions to the potential are $\sim N_c$ and the subleading ones
are $\sim 1/N_c$, the neglected terms are of relative $1/N_c^2$ order,
so we might expect an {\it a priori} $\sim 10\%$, accuracy.

The coincidence between $^1S_0$ and $^3S_1$ potentials complies to the
Wigner $SU(4)$ symmetry which we review for completeness in
Appendix~\ref{eq:su4} for the two-nucleon system.  Modern high quality
potentials~\cite{Stoks:1994wp} describing accurately NN scattering
below pion production threshold show some traces of the symmetry for
distances above $1.4-1.8 {\rm fm}$.  Quenched lattice QCD evaluations
of NN potentials for $m_\pi / m_\rho \sim
0.6$~\cite{Ishii:2006ec,Aoki:2008hh} yield also similar $^1S_0$ and
$^3S_1$ potentials for $r > 1.4 {\rm fm}$. Thus, at first sight one
may conclude that Wigner symmetry holds when OPE dominates, and thus
has a limited range of applicability. An important result of the
present investigation, which will be elaborated along the paper, is
that this is not necessarily so, provided the relevant scales of
symmetry breaking are properly isolated with the help of
renormalization ideas.

Let us analyze the consequences of the
symmetry, Eq.~(\ref{eq:pot}), within the standard approach to OBE
potentials.  The scattering phase-shift $\delta_0(p)$ is computed by
solving the (S-wave) Schr\"odinger equation in r-space
\begin{eqnarray}
-u''_p(r) + M_{N}\,V(r)\,u_p(r) &=& p^2\,u_p(r) \, ,\label{eq:Scrod-p} \\ 
u_p(r) &\to& \frac{\sin{\left(p r + \delta_0 (p)\right)}}{\sin{\delta_0(p)}}
\, , 
\label{eq:up-asymp}
\end{eqnarray}
with a regular boundary condition at the origin
$u_p(0)=0$. 
Moreover, for a short range potential such as
the one in Eq.~\eqref{eq:pot} one also has the Effective Range
Expansion (ERE)
\begin{eqnarray}
p \cot{\delta_0 (p)} = -\frac{1}{\alpha_0} + \frac{1}{2}\,r_0\,p^2 +
\cdots \, ,
\label{eq:ere}
\end{eqnarray}
where the \textit{scattering length}, $\alpha_0$, is 
defined by the asymptotic behavior
of the zero energy wave function as 
\begin{eqnarray} 
u_0 (r) &\to & 1- \frac{r}{\alpha_0}  \, ,  
\end{eqnarray} 
and the effective range, $r_0$, is given by 
\begin{eqnarray} 
r_0 &=& 2 \int_0^\infty dr \left[ \left(1-\frac{r}{\alpha_0} \right)^2-
u_0 (r)^2 \right] \, . 
\label{eq:r0_singlet}
\end{eqnarray} 
In the usual approach~\cite{Machleidt:1987hj,Machleidt:2000ge}
everything is obtained from the potential assumed to be valid for
$0\leq r < \infty$. We note incidentally that the Wigner symmetry
relation, Eq.~(\ref{eq:pot}), holds at {\it all}
distances~\footnote{In practice, strong form factors are included
mimicking the finite nucleon size and reducing the short distance
repulsion of the potential, but the regular boundary condition is
always kept.}.  In addition, due to the \textit{unnaturally large} NN
$^1S_0$ scattering length ($\alpha_s \sim -23 {\rm fm}$), any change
in the potential $V \to V + \Delta V$ has a dramatic effect on
$\alpha_0$, since one obtains
\begin{eqnarray}
\Delta\alpha_0 = \alpha_0^2 M_N \int_0^{\infty} \Delta V(r)
u_0(r)^2\mathrm{d}r \, ,
\end{eqnarray}
and thus the potential parameters \textit{must be fine tuned}, and in
particular the short distance physics. As it was discussed in
Refs.~\cite{RuizArriola:2007wm,CalleCordon:2008eu} this short distance
sensitivity is unnatural as long as the OBE potential does not truely
represent a fundamental NN force at short distances. Indeed, the
sensitivity manifests itself as tight constraints for the potential
parameters when the $^1S_0$ phase shift is fitted resulting in
incompatible values of the coupling constants as obtained from other
sources as NN scattering. Of course, there is nothing wrong in the
need of a fine tuning as this is a unavoidable consequence of the
large scattering length; the relevant point is whether this should be
driven by a potential which will not be realistic at short distances.

In any case, in the traditional approach to NN potentials we are
confronted with a paradox; on the one hand the symmetry seems to
suggest that {\it somewhere} the phase shifts should coincide, while
on the other hand a fine tuning is required because of the large
scattering lengths.  In the standard approach, if $V_s (r)= V_t(r) $
then $\delta_s(k)=\delta_t(k)$ and thus $\alpha_s=\alpha_t$, as one
naturally expects. A straightforward explanation, of course, is to
admit that the symmetry is strongly violated. This would make
difficult to understand how can $SU(4)$ work at all for light nuclei
if the simpler two nucleon system does not show manifestly the
symmetry.

Before presenting our solution to this dilemma in the next section,
let us note that a good condition for an approximate symmetry is that
it be stable under symmetry breaking, otherwise a tiny perturbation $
V_s(r) - V_t(r) =\Delta V (r) \neq 0 $ would yield a large change, and
this is precisely the bizarre situation we are bound to evolve because
of the large scattering lengths. This suggests a clue to the problem,
namely we should provide a framework where the highly
potential-sensitive scattering length becomes a variable independent
of the potential. More generally, we want to avoid the logical
conclusion that a symmetry of the potential is a symmetry of the
S-matrix~\footnote{This situation resembles the case of anomalies in
Quantum Field Theory where the parallel statement would be that a
symmetry of the Lagrangean becomes a symmetry of the S-matrix, a
conclusion which may be invalidated by the impossibility of preserving
the symmetry by the necessary regularization of loop integrals. The
present case is a bit more subtle as it corresponds to the case of
finite but ambiguous theories (see e.g. Ref.~\cite{Jackiw:1999qq}) }.
As we will explain below the puzzle may be overcome by the concept of
long distance symmetry; a symmetry which is only broken at short
distances by a suitable boundary condition.

\section{Universality relations and Renormalization}
\label{sec:uni-renor}

We cut the Gordian knot by appealing to renormalization in coordinate
space~\cite{PavonValderrama:2005wv,PavonValderrama:2007nu}. As we will
show this enables to disentangle short and long distances in a way
that the symmetry is kept at all non-vanishing distances. The main
idea is to fix the scattering length independently of the potential by
means of a suitable short distance boundary condition. As a result the
undesirable dependence of observables on the potential is reduced at
short distances, precisely the region where a determination of the NN
force in terms of hadronic degrees of freedom becomes less reliable.

Let us review in the case of S-waves the renormalization
procedure in coordinate space pursued
elsewhere~\cite{PavonValderrama:2005wv,PavonValderrama:2007nu} and
which will prove particularly suitable in the sequel. This is fully
equivalent to introduce one counter-term in the cut-off
Lippmann-Schwinger equation in momentum space (see
Ref.~\cite{Entem:2007jg} for a detailed discussion on the
connection). The superposition principle of boundary conditions
implies,
\begin{eqnarray}
u_k (r) = u_{k,c} (r)  +  k \cot \delta_0 \,  u_{k,s} (r)  \, ,   
\label{eq:sup_k}
\end{eqnarray} 
with $ u_{k,c} (r) \to \cos (k r) $ and $ u_{k,s} (r) \to \sin (k r)
/k $ for $r \to \infty $. At zero energy, $k \to 0$, and $\delta_0 (k) \to -
\alpha_0 k $ yields 
\begin{eqnarray}
u_0 (r) = u_{0,c} (r)  - \frac1{\alpha_0}  u_{0,s} (r)    \, ,
\label{eq:sup_0}
\end{eqnarray} 
with $ u_{0,c} (r) \to 1 $ and $ u_{0,s} (r) \to  r $ for $r \to
\infty $. Combining the zero and finite energy wave functions we get
\begin{eqnarray}
\left[u_k'(r) u_0 (r) - u_0'(r) u_k(r) \right] \Big|_{r_c}^\infty = k^2
\int_{r_c}^\infty u_k(r) u_0(r) dr \, , \nonumber \\ 
\end{eqnarray} 
where $r_c$ is a short distance cut-off radius which will be removed
at the end. To calculate the contribution from the term at infinity
we use the long distance behavior, Eq.~(\ref{eq:up-asymp}). The
integral and the boundary term at infinity yield two canceling delta
functions.  This corresponds to take
\begin{eqnarray}
\int_0^\infty u_k(r) u_p(r) dr = \frac{\pi \delta (k-p) }{2 \sin^2
\delta_0 (k)} \, , 
\label{eq:norm}
\end{eqnarray} 
as can be readily seen.  We are thus left with the boundary term at
short distances, taking the limit $r_c \to 0$ we get
\begin{eqnarray}
\lim_{r_c \to 0} \left[ u_k'(r_c) u_0(r_c) - u_0'(r_c) u_k(r_c)
\right]=0 \, .
\end{eqnarray} 
Note that the regular solution $u_k(r_c)=u_0(r_c)=0$ is a {\it
particular} choice for $r_c=0$. Writing out the orthogonality
condition via the superposition principle at finite and zero energies,
Eq.~(\ref{eq:sup_k}) and Eq.~(\ref{eq:sup_0}) respectively, one gets
\begin{eqnarray}
0= \int_0^\infty dr && \left[u_{0,c} (r) - \frac{1}{\alpha_0} \,
u_{0,s} (r) \right] \nonumber \\ & & \times \Big[ u_{k,c} (r) + k \cot
\delta_0 (k) \, u_{k,s} (r) \Big] \, .
\end{eqnarray} 
Expanding the integrand and defining 
\begin{eqnarray}
{\cal A}(k) &=& \int_0^\infty dr \, u_{0,c} (r) u_{k,c} (r) \, ,
\nonumber \\ {\cal B}(k) &=& \int_0^\infty dr \, u_{0,s} (r) u_{k,c}
(r) \, , \nonumber \\ {\cal C}(k) &=& \int_0^\infty dr \, u_{0,c} (r)
u_{k,c} (r) \, , \nonumber \\ {\cal D}(k) &=& \int_0^\infty dr \,
u_{0,s} (r) u_{k,s} (r) \, ,
\label{eq:ABCD} 
\end{eqnarray} 
we get the explicit formula
\begin{eqnarray}
k \cot \delta_0 (k) = \frac{ \alpha_0 {\cal A} ( k) + {\cal B} (k)}{
\alpha_0 {\cal C} ( k) + {\cal D} (k)} \, . 
\label{eq:phase_singlet}
\end{eqnarray} 
The functions ${\cal A}$, ${\cal B}$, ${\cal C}$ and ${\cal D}$ are
even functions of $k$ which depend {\it only on the potential}. Note
that the dependence of the phase-shift on the scattering length is
wholly {\it explicit}; $\cot \delta_0 $ is a bilinear rational mapping
of $\alpha_0$. Further, using Eq.~(\ref{eq:sup_0}), one gets the
effective range 
\begin{eqnarray} 
r_0  &=&  A + \frac{B}{\alpha_0}+ \frac{C}{\alpha_0^2}  \, ,    
\label{eq:r0_univ} 
\end{eqnarray} 
where 
\begin{eqnarray}
A &=& 2 \int_0^\infty dr ( 1 - u_{0,c}^2 ) \, , \\    
B &=& -4 \int_0^\infty dr ( r - u_{0,c} u_{0,s} ) \, , \\    
C &=& 2 \int_0^\infty dr ( r^2 - u_{0,s}^2 )    \, ,  
\end{eqnarray} 
depend on the potential parameters only. Again, the interesting thing
is that all explicit dependence on the scattering length $\alpha_0$ is
displayed by Eq.~(\ref{eq:r0_univ} ).

We turn now to discuss the case of a bound state corresponding to the
case of negative energy $E=-\gamma^2/M$ where $\gamma$ is the
wave number. The wave function behaves asymptotically as  
\begin{eqnarray}
u_\gamma (r) \to A_S e^{-\gamma r} \, ,  
\label{eq:bound-long}
\end{eqnarray} 
and is chosen to fulfill the normalization condition
\begin{eqnarray}
\int_0^\infty u_\gamma (r)^2 dr=1 \, . 
\end{eqnarray} 
In principle, such a state would be unrelated to the scattering
solutions. An explicit relation may be determined from the
orthogonality condition, which applied in particular to the zero
energy state yields
\begin{eqnarray}
0= \int_0^\infty dr \left[u_{0,c} (r) - \frac{1}{\alpha_0} 
u_{0,s} (r) \right] u_\gamma (r) \, . 
\end{eqnarray} 
This generates a correlation between the scattering length, $\alpha_0$
and the bound state wave number, $\gamma$,
\begin{eqnarray}
\alpha_0 (\gamma) = \frac{\int_0^\infty dr u_\gamma(r) u_{0,s} (r)}
{\int_0^\infty dr u_\gamma (r) u_{0,c} (r) } \, . 
\end{eqnarray} 
We remind that the two independent zero energy solutions, $u_{0,c}
(r)$ and $u_{0,s} (r)$ depend only on the potential.

A trivial realization of the conditions discussed above is given by
the case where there is no potential, $U(r)=0$. Hence, the general
solution for a positive energy state $E=k^2/M$ is given by
\begin{eqnarray}
u_k (r) = \cot \delta_0 (k) \sin (kr) + \cos(kr)  \, ,  
\end{eqnarray} 
and using the low energy limit condition $\delta_0 (k) \to - \alpha_0
k $ we obtain 
\begin{eqnarray}
u_0 (r) = 1 - \frac{r}{\alpha_0}  \, . 
\end{eqnarray} 
Orthogonality between zero and finite energy states yields after
evaluating the integrals 
\begin{eqnarray}
k \cot \delta_0 (k) = - \frac1{\alpha_0} \, , 
\end{eqnarray} 
and as a consequence the effective range vanishes $r_0=0$, in
accordance to the fact that the range of the potential is zero. For a
negative energy state $E=-\gamma^2/M$ the normalized bound state is
\begin{eqnarray}
u_\gamma (r) = A_S e^{-\gamma r}\, , \qquad A_S = 1/\sqrt{2 \gamma}\, . 
\end{eqnarray} 
Orthogonality between the zero energy and the bound state, again,
yields the correlation 
\begin{eqnarray}
\alpha_0 = 1/\gamma \, .  
\end{eqnarray} 
In the appendix~\ref{sec:pert} we illustrate further the procedure in
the case of weak potentials for which a form of perturbation theory
may be applied for the case of {\it weak potentials} but {\it
arbitrary} scattering lengths.

Before going further we should ponder on the need to take the limit
$r_c \to 0$, which corresponds to eliminating the cut-off. We note
that the potential, $V(r)$, is used at {\it all} distances both in the
standard approach, which involves the regular solution only, and the
renormalization approach, which requires the regular as well as the
irregular solution.  However, the sensitivity to the short distance
behaviour of the potential is quite different; the standard approach
displays much stronger dependence while the renormalization approach
is fairly independent on the hardly accessible short distance region,
a feature that becomes evident perturbatively (see
e.g. Eq.~(\ref{eq:r0_pert})).  This is in fact the key property that
allows to eliminate the cut-off in the renormalization approach.
Thus, removing the cut-off does not mean that the OBE potential is
believed to hold all the way down to the origin.

The procedure carried out before is described in purely quantum
mechanical terms, but it can be mapped onto field theoretical
terminology; it is equivalent to the method of introducing one
counter-term in the cut-off Lippmann-Schwinger equation in momentum
space~\cite{Entem:2007jg,Valderrama:2007ja}. Moreover,
Eq.~(\ref{eq:sup_0}) represents the corresponding renormalization
condition, which is chosen to be on-shell at zero energy. In the case
of the bound state the corresponding renormalization condition is
given by Eq.~(\ref{eq:bound-long}) at negative energy. Imposing more
than one renormalization condition, i.e. introducing more than one
counter-term and removing the cut-off presents some subtleties which
have been discussed in Refs.~~\cite{PavonValderrama:2007nu,
Entem:2007jg}. We will analyze below this issue in the present context
(see Sect.~\ref{sec:two-ct}).

\begin{figure}
\includegraphics[height=.36\textheight,angle=270]{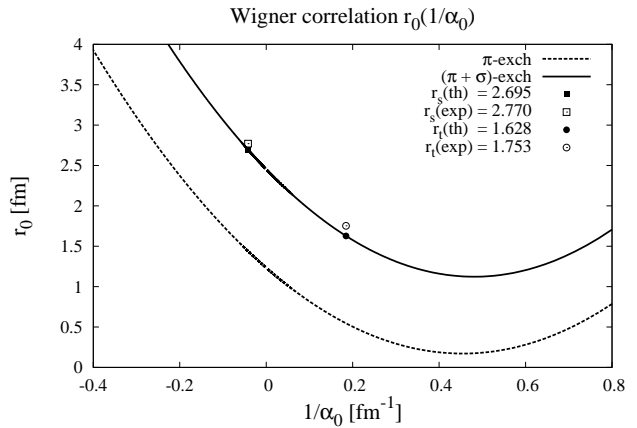}
\caption{The Wigner correlation for the effective range $r_0 $ (in fm)
of a np S-wave as a function of an arbitrary inverse scattering length
$\alpha_0$ in the case of the OPE and OPE+$\sigma$ potentials.  The
parabolic shape is determined by a unique long distance potential. The
points in the solid curve correspond to the two different values of
the effective range $r_s$ in the singlet $^1S_0$ and $r_t$ in the
triplet $^3S_1$ channels when the scattering length is taken to be
$\alpha_s = - 23.74 {\rm fm}$ and $\alpha_t = 5.42 {\rm fm}$
respectively. Experimental points are also shown for comparison.}
\label{fig:r0-a0}
\end{figure}

\section{Central phases and the deuteron}
\label{sec:numeric}

\subsection{Potential Parameters}

To proceed further we fix the potential parameters, keeping in mind
that the leading $N_c$ nature of the potential embodies some
systematic $1/N_c^2$ uncertainties. Of course, while we will use
relations which are compatible with large $N_c$ scaling, the numerical
values can only be fixed phenomenologically.  The main point is that
besides the $\sigma$-meson mass (see below), we may choose quite
natural values for the masses and couplings unlike the usual OBE
potentials~\cite{Machleidt:1987hj}. As was discussed at the end of
Sect.~\ref{sec:OBE} the standard approach suffers from tight
constraints reflecting the unnatural short distance sensitivity.  In
this regard, let us note that, as emphasized in
Refs.~\cite{RuizArriola:2007wm,CalleCordon:2008eu}, it is a virtue of
the renormalization viewpoint which we are applying here to the OBE
potential, that the unwanted short distance sensitivity is largely
removed, allowing for a determination of the potential parameters
using independent sources. For definiteness we take $g_{\pi NN}=13.1$
and $g_{\sigma NN}=10.1$, quite close to the Goldberger-Treiman values
for $\sigma$ and $\pi$, $g_{\sigma NN} = M_N /f_\pi $ and $g_{\pi NN}
= g_A M_N /f_\pi $ respectively.  We also take the SU(3) value
$g_{\omega NN} = 3 g_{\rho NN} - g_{\phi NN} $ which on the basis of
the OZI rule, $g_{\phi NN} =0$, Sakurai's universality $ g_{\rho NN} =
g_{\rho \pi \pi} /2$ and the KSFR relation $ 2 g_{\rho \pi \pi}^2
f_\pi^2 = m_\rho^2 $ yields $g_{\omega NN} = N_c m_\rho /( 2 \sqrt{2}
f_\pi)= 8.8$. The rho tensor coupling is taken to be $f_{\rho NN}=
\sqrt{2} M_N g_{\omega NN} /m_\rho = 15.5 $ which cancels the vector
meson contributions in the potential and yields $\kappa_\rho = f_{\rho
NN} /g_{\rho NN} = 5.5$ a quite reasonable
result~\cite{Machleidt:1987hj}~\footnote{As shown in previous
work~\cite{RuizArriola:2007wm,CalleCordon:2008eu} the net vector meson
exchange contribution corresponding to the combined repulsive coupling
$g_{\omega NN}^2 - f_{\rho NN}^2 m_\rho^2 / 2 M_N^2 $ (referred there
simply as $g_{\omega NN}^2$) cannot be pinned down accurately from a
fit to the $^1S_0$ phase shift being compatible with zero within
errors. This is due to the short distance insensitivity embodied by
the renormalization approach.}. Note that $1/N_c$ effects include not
only other mesons but also finite width effects of $\sigma$ and $\rho$
since for large $N_c$ one has stable mesons, $\Gamma_\sigma,
\Gamma_\rho \sim 1/N_c$.  For the masses we take $m_\pi=140 {\rm MeV}$
and $m_\omega=783{\rm MeV}$. This fixes all parameters except
$m_\sigma$ (actually the real part) which we identify with the
lightest $J^{PC}=0^{++}$ meson $f_0(600)$. According to the recent
analysis based on Roy equations $m_\sigma - i \Gamma_\sigma /2 =
441^{+16}_{-8} - i 272^{+9}_{-12} {\rm MeV}$~\cite{Caprini:2005zr}. A
fit to the pn data of Ref.~\cite{Stoks:1994wp} in the $^1S_0$ channel
yields $m_\sigma=510(1) {\rm MeV}$, where the error is
statistical. The fitted mass value differs by about $10\%$ from the
location of the real part of the resonance, in harmony with the
expected $1/N_c^2$ corrections~\footnote{Actually, our estimate of the
$\sigma$-mass as a pole in the second Riemann sheet for $\pi\pi$
scattering for large $N_c$~\cite{CalleCordon:2008eu} yields the value
$m_\sigma \sim 507 {\rm MeV}$.}. Although a more quantitative estimate
of the large $N_c$ corrections to the potentials parameters would be
very useful, for the present purposes of discussing Wigner symmetry on
the light of large $N_c$ it is more than sufficient. Thus, we make no
attempt here to make any systematic expansion.

\begin{figure}[tbc]
\includegraphics[height=4cm,width=4.5cm,angle=270]{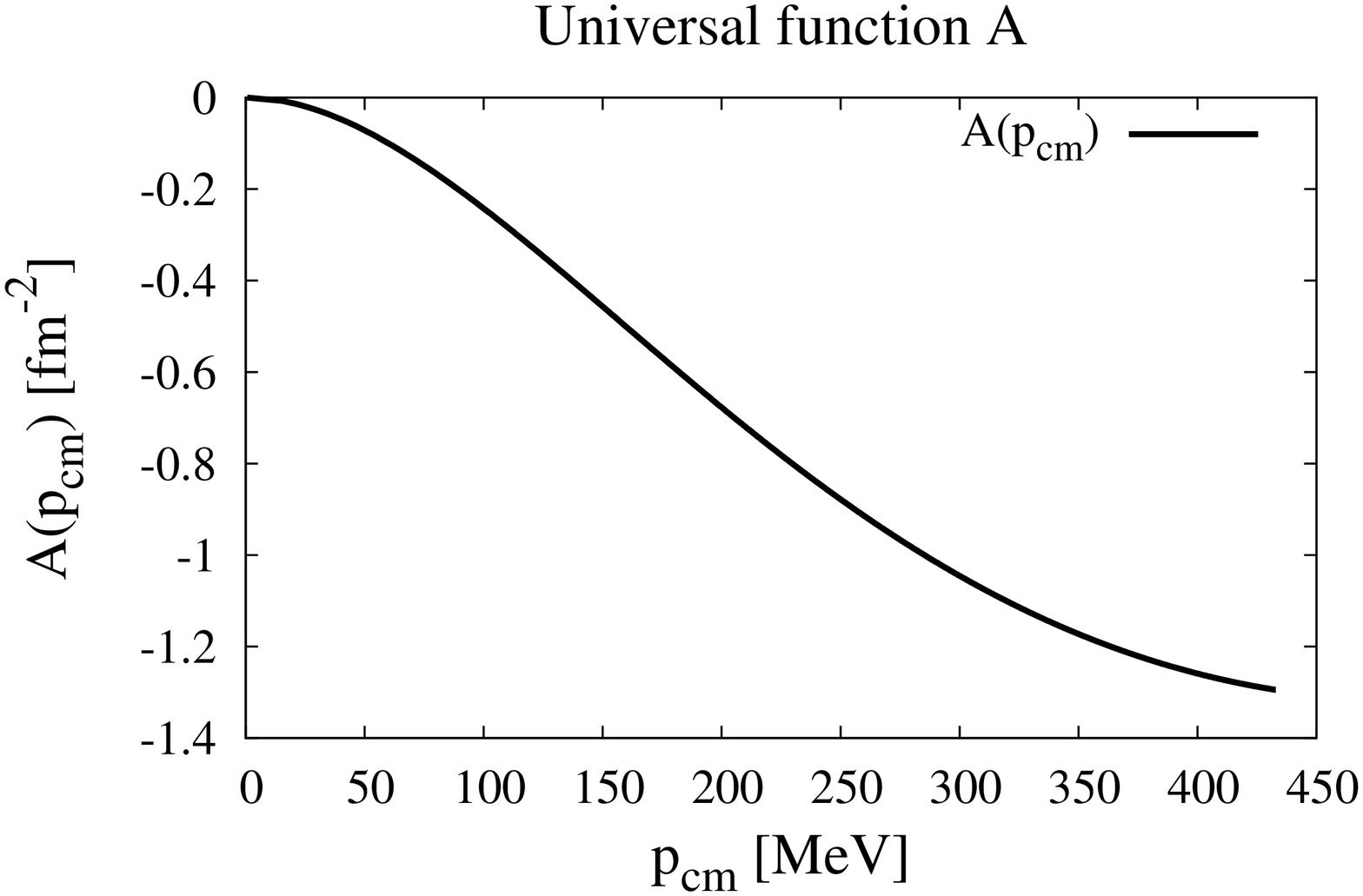}
\includegraphics[height=4cm,width=4.5cm,angle=270]{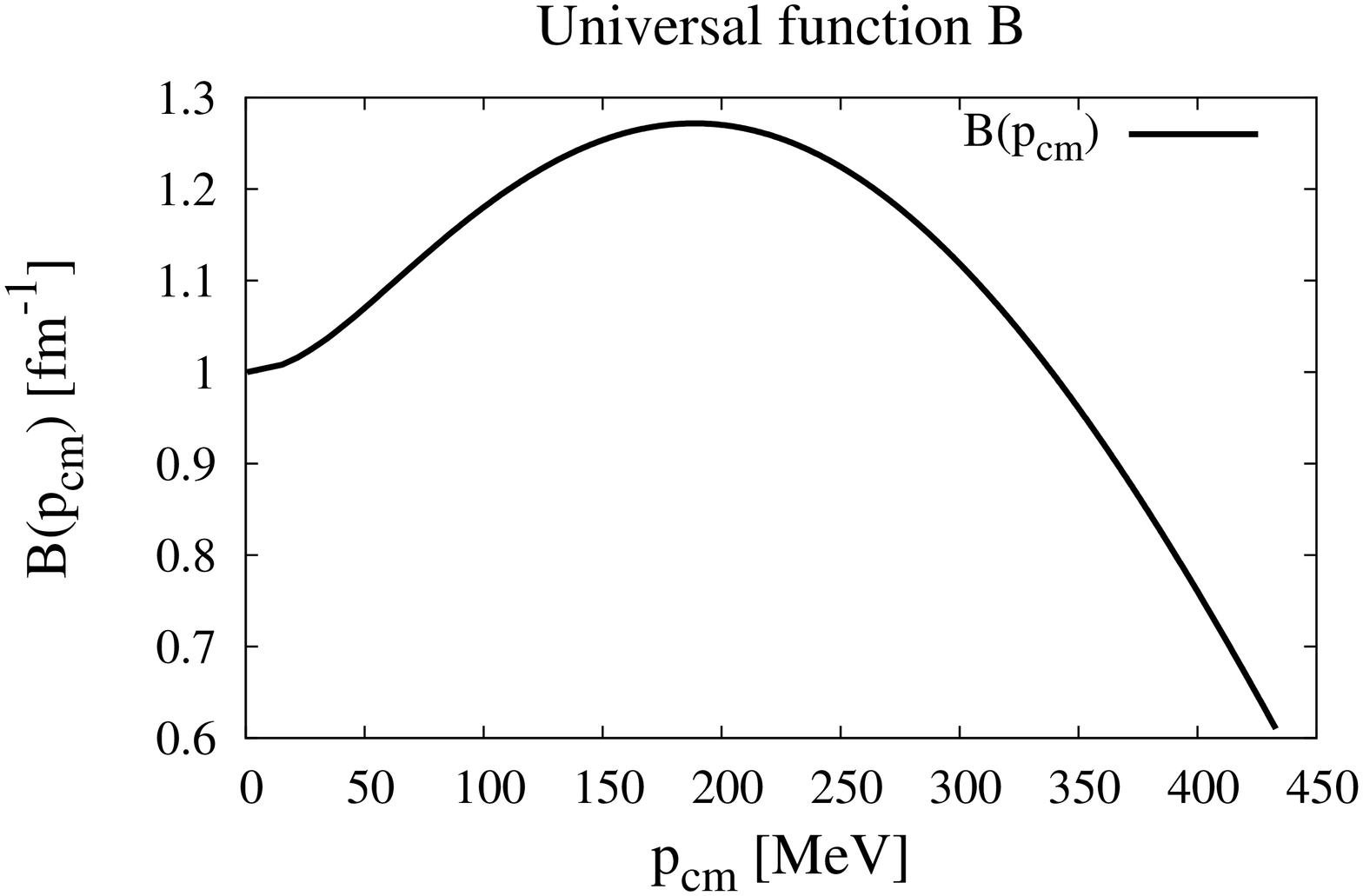} \\
\includegraphics[height=4cm,width=4.5cm,angle=270]{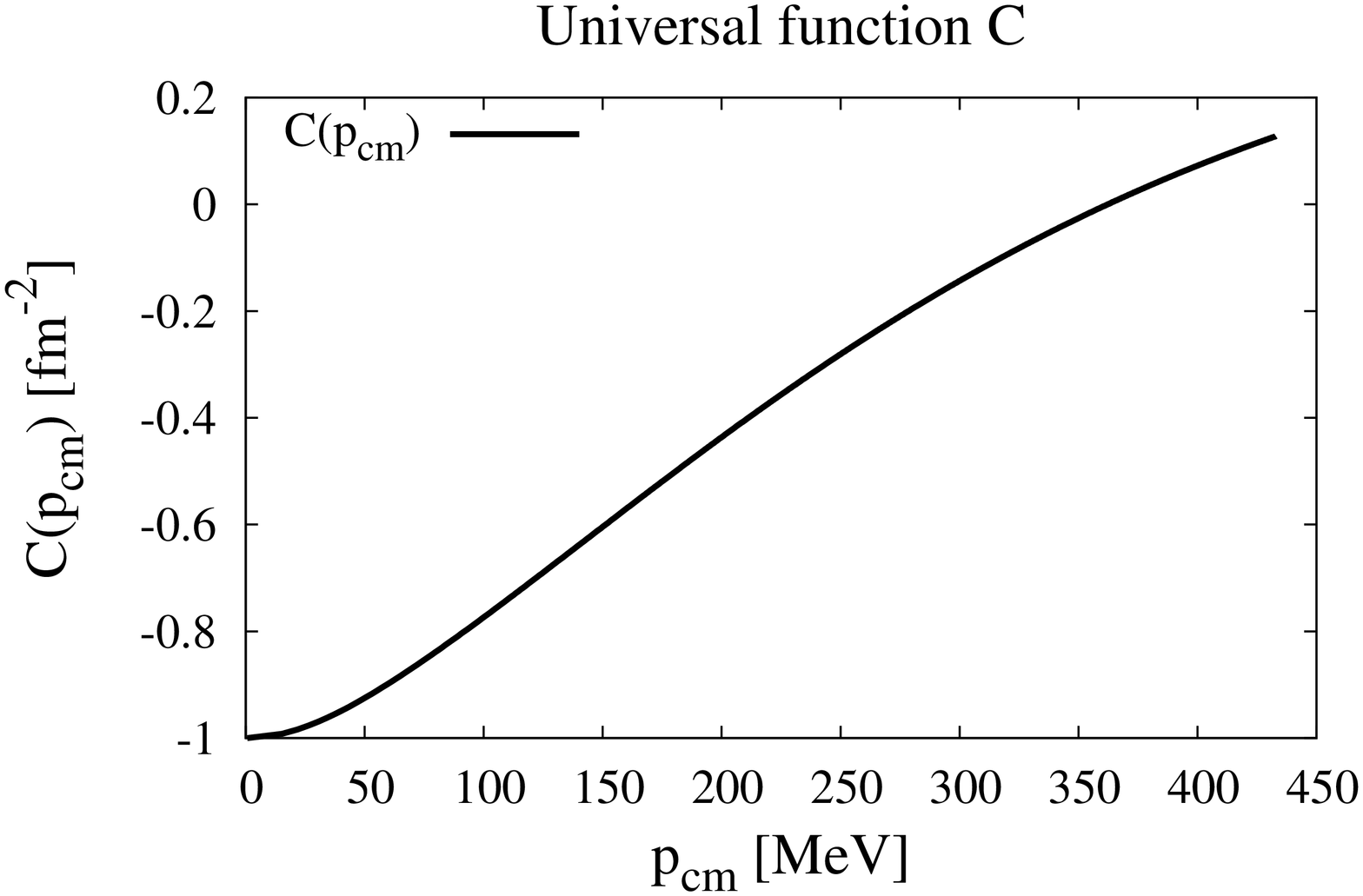}
\includegraphics[height=4cm,width=4.5cm,angle=270]{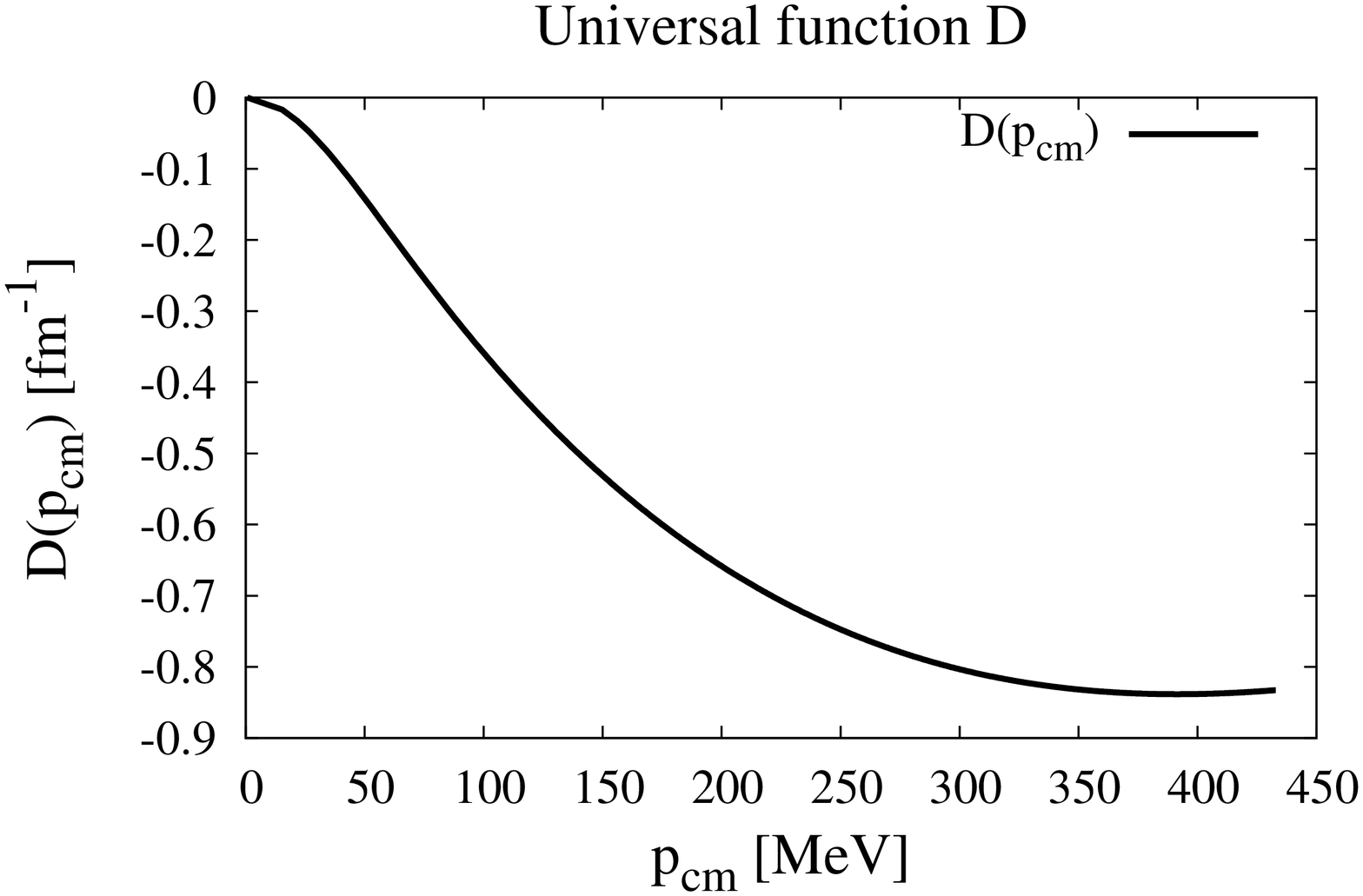}
\caption{The universal functions ${\cal A}$, ${\cal B}$ ${\cal C}$ and
${\cal D}$ defined by Eqs.~(\ref{eq:ABCD}) in appropriate length units
as a function of the CM momentum $p$ (in MeV). These functions depend
on the potential $V_s (r) = V_t(r) $ only but are independent of the
scattering length.}
\label{fig:ABCD}
\end{figure}

\subsection{Low energy parameters and phase shifts}

Clearly, in the traditional approach if we have $V_s(r)=V_t(r)$ and
impose the regular boundary condition, $u_s(0)=u_t(0)=0$, the only
possible solution is $\alpha_s=\alpha_t$, $r_s=r_t$ and $\delta_s
(p)=\delta_t (p)$. However, in the renormalization approach we allow
{\it different} short distance boundary conditions $u_s' (0^+) / u_s
(0^+) \neq u_t' (0^+) / u_t (0^+)$~\footnote{The limit from above, $u
(0^+) = \lim_{r_c \to 0^+} u(r_c)$ is really necessary to pick both
the irregular and irregular solutions. If one starts {\it exactly}
from the origin the only possible solution is the regular one.}, and
hence we may have $\alpha_s \neq \alpha_t$. Note that this corresponds
to a breaking of the symmetry at {\it short} distances and hence
postulating its validity at {\it long} distances. The previous
equations imply straight away the following expressions for the
effective ranges in the singlet and triplet channels, 
\begin{eqnarray} 
r_s  &=&  A + \frac{B}{\alpha_s}+ \frac{C}{\alpha_s^2}  \, ,  \nonumber \\   
r_t  &=&  A + \frac{B}{\alpha_t}+ \frac{C}{\alpha_t^2}  \, .     
\label{eq:r0_univ_st} 
\end{eqnarray} 
As already mentioned, the remarkable aspect of these two equations is
the fact that the coefficients $A,B,C$ are {\it identical} both in the
triplet as well as in the singlet channels as long as $V_s(r)=V_t(r)$,
thus the only difference resides in the numerical values of the
scattering lengths $\alpha_s$ and $\alpha_t$. Numerically we get
(everything in fm )
\begin{eqnarray}
r_0 &=& 1.3081 -
\frac{4.5477}{\alpha_0}+\frac{5.1926}{\alpha_0^2} \, \qquad (\pi)
\nonumber 
\\ 
&=& 1.5089 {\rm fm} \, \quad ( \alpha_0 = \alpha_s) \quad  
({\rm exp.} 2.770 {\rm fm})   
\nonumber 
\\
&=& 0.6458 {\rm fm} \, \quad ( \alpha_0 = \alpha_t) \quad  
({\rm exp.} 1.753 {\rm fm})   
\nonumber 
\\
r_0&=& 2.4567 - \frac{5.5284}{\alpha_0} + \frac{5.7398}{\alpha_0^2}
\qquad (\pi+\sigma) \nonumber \\
&=& 2.6989 {\rm fm} \, \quad ( \alpha_0 = \alpha_s) \quad 
({\rm exp.} 2.770 {\rm fm})      
\nonumber 
\\
&=& 1.5221 {\rm fm} \, \quad ( \alpha_0 = \alpha_t) 
\quad  
({\rm exp.} 1.753 {\rm fm})
\label{eq:r0_univ_st_num}       
\end{eqnarray} 
where the corresponding numerical values when the experimental
$\alpha_s = -23.74 {\rm fm} $ and $\alpha_t = 5.42 {\rm fm} $ as well
as the experimental values for the effective ranges have also been
added. More generally, for any fixed potential the correlation of
$r_0$ on $1/\alpha_0$ is a parabola which we plot in
Fig.~\ref{fig:r0-a0} for the OPE and OPE+$\sigma$. This dependence is
universal to {\it all} S-waves having the {\it same} potential and
from this viewpoint there is nothing in this curve making unnaturally
large scattering lengths particularly different from smaller ones. The
present analysis, however, does not shed any light on the origin of
the large size of the $\alpha's$ nor {\it how} $\alpha_s$ and
$\alpha_t$ are interrelated~\footnote{This is in fact a price we pay
for the built-in short distance insensitivity. We note, however, that
after Refs.~\cite{Epelbaum:2002gb,Beane:2002xf,
Braaten:2003eu,Epelbaum:2006jc,Hammer:2007kq} both scattering lengths
might coincide for a pion mass around $m_\pi \sim 200 {\rm MeV}$.  As
a consequence, QCD might be close to a point where the effective
theory had a standard $SU(4)$ symmetry at {\it zero} energy. Actually,
in Ref.~\cite{Beane:2002xf} the similarity between $^1S_0$ and $^3S_1$
phase shifts can be seen. This scenario would turn the long distance
symmetry we propose for the physical pion mass into a standard
symmetry for such an unphysical value of the pion mass.}. In any case,
as we see from Fig.~\ref{fig:r0-a0}, the experimental values fall
strikingly almost on top of the curve, pointing towards a correct
interpretation of the underlying symmetry.

\begin{figure}[tbc]
\includegraphics[height=8cm,width=5.5cm,angle=270]{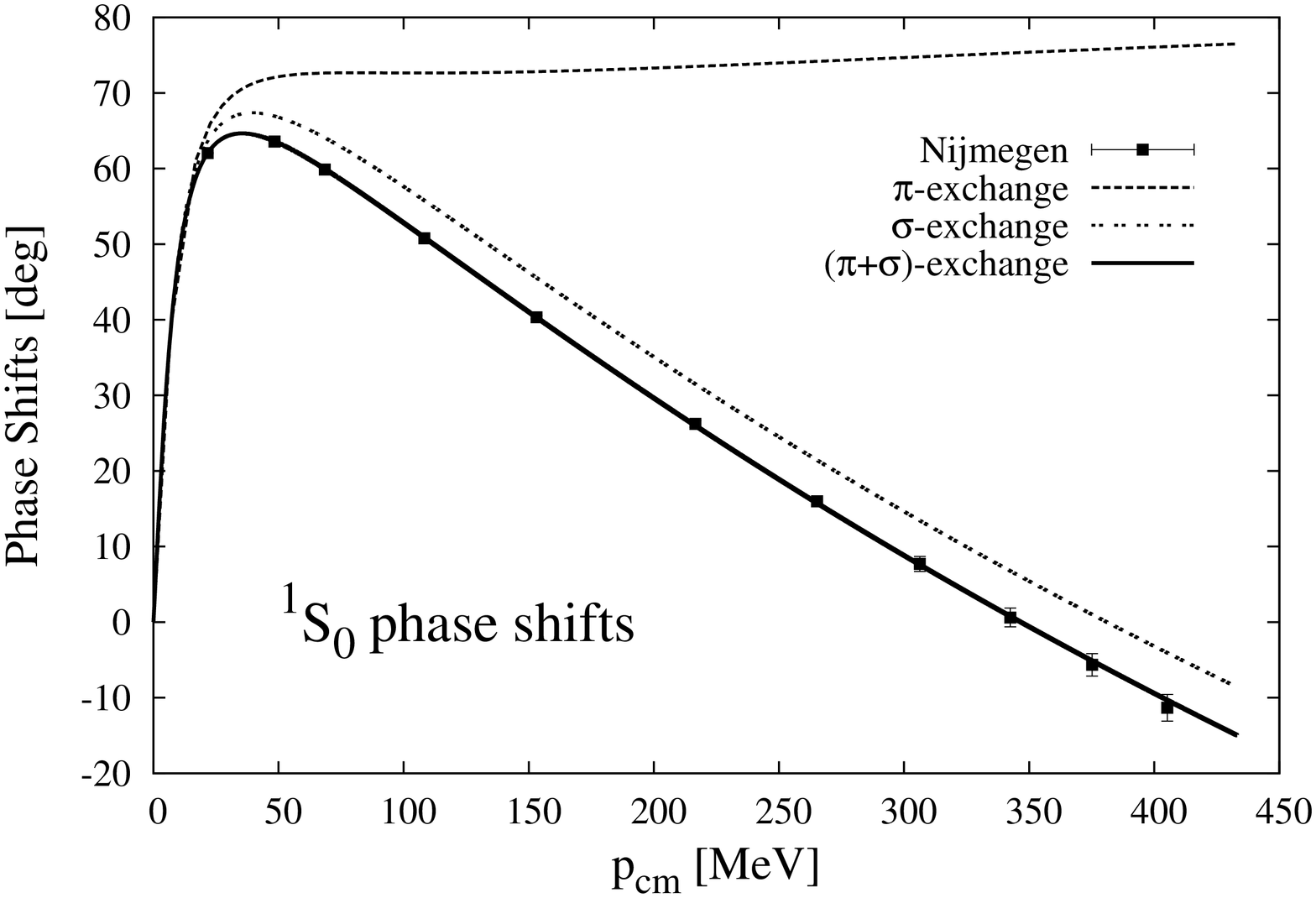}\\
\includegraphics[height=8cm,width=5.5cm,angle=270]{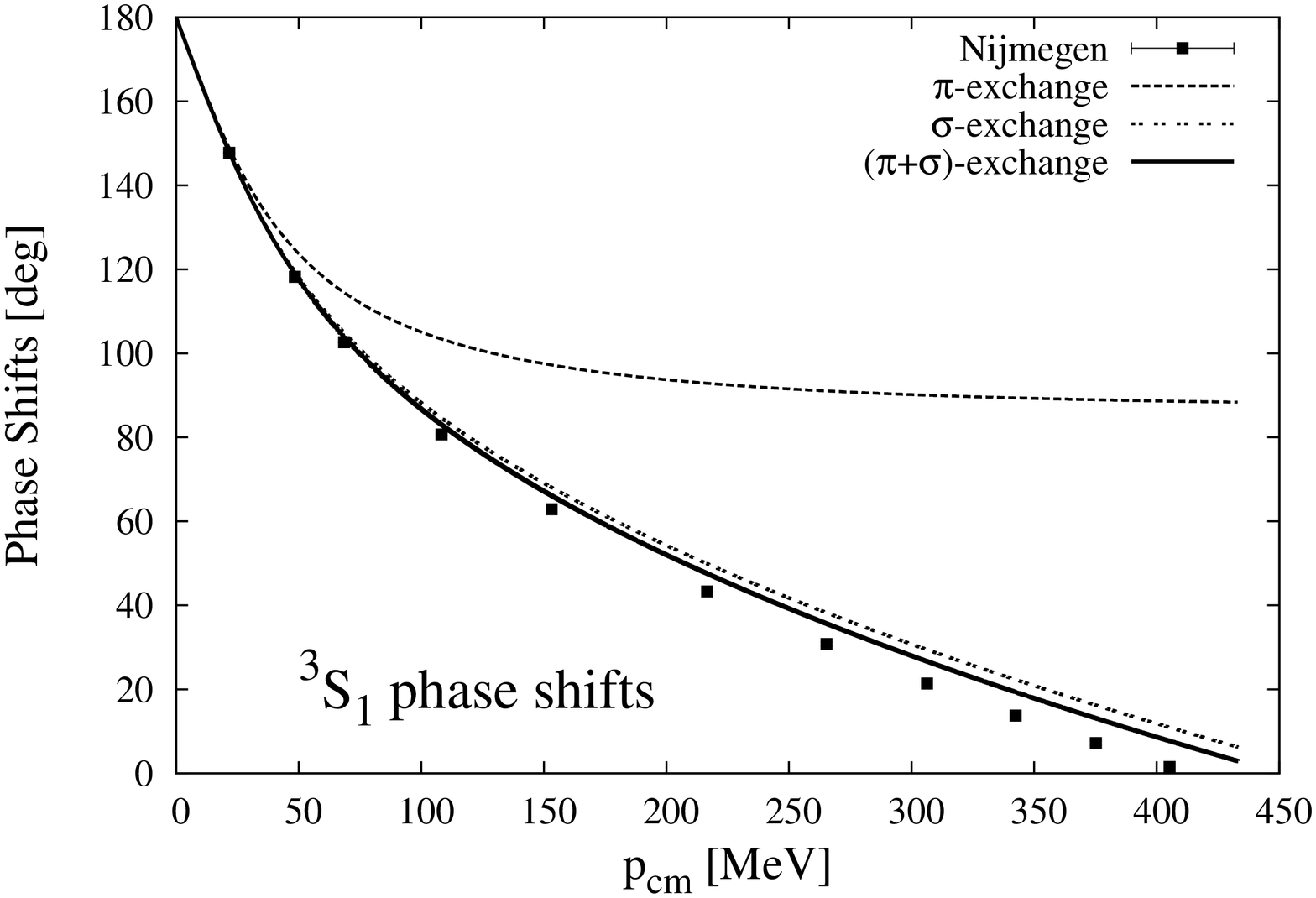}
\caption{Phase shifts (in degrees) for the {\it fitted} $^1S_0$ (top
panel) and {\it predicted} $^3S_1$ (bottom panel) channels as a
function of the CM momentum (in MeV). In both cases the potential is
the {\it same}, $V_s(r)=V_t(r)$, while the only difference is in the
scattering lengths in the singlet channel $\alpha_s = - 23.74 {\rm
fm}$ and in the triplet channel $\alpha_t = 5.42 {\rm fm}$
corresponding to a {\it different} short distance boundary
condition. We also plot the cases with only $1\sigma$-exchange and
$1\pi$-exchange for comparison.  Data from \cite{Stoks:1994wp}.}
\label{fig:singlet-triplet}
\end{figure}

We turn next to the phase shifts.  According to
Eq.~(\ref{eq:phase_singlet}) they are given in terms of the universal
functions ${\cal A}$, ${\cal B}$ ${\cal C}$ and ${\cal D}$ defined by
Eqs.~(\ref{eq:ABCD}) and presented in Fig.~\ref{fig:ABCD} in
appropriate length units as a function of the CM momentum $p$ in MeV
for completeness.  As we see, these functions are smooth.
From them the corresponding singlet and triplet phase shifts are obtained by 
\begin{eqnarray}
k \cot \delta_s &=& \frac{ \alpha_s {\cal A} ( k) + {\cal B} (k)}{
\alpha_s {\cal C} ( k) + {\cal D} (k)} \nonumber \\ k \cot \delta_t
&=& \frac{ \alpha_t {\cal A} ( k) + {\cal B} (k)}{ \alpha_t {\cal C} (
k) + {\cal D} (k)} \, , 
\end{eqnarray} 
respectively. When the experimental scattering lengths $\alpha_s = -
23.74 {\rm fm}$ and $\alpha_t = 5.42 {\rm fm}$ are taken we can {\it
  fit} the singlet $^1S_0$ channel and {\it predict} the triplet
$^3S_1$ channel. 

The result is shown in Fig.~\ref{fig:singlet-triplet} and as we see
the agreement is remarkably good taking into account that we have
neglected the tensor force and the {\it a priori} $1/N_c^2$ systematic
corrections to the potential. Note that the identity of the singlet
and triplet potentials is not sufficient; the simple OPE fulfills this
property but does not explain the neither phase-shifts. Actually, it
shows that {\it both} failures are correlated~\footnote{The reason why
OPE fails at much lower energies in the $^1S_0$ channel than in the
$^3S_1$ channel is due to a stronger short distance sensitivity of the
channel with larger scattering length.}.

\subsection{Renormalization and scale invariance}

\begin{figure}[tbc]
\includegraphics[height=.36\textheight,angle=270]{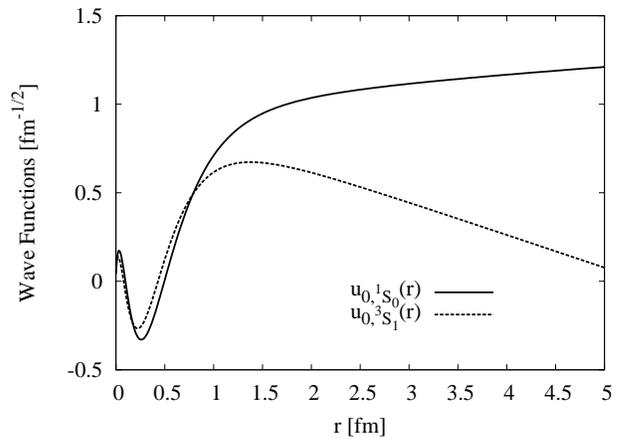}
\caption{Zero energy S-waveradial functions for the singlet $^1S_0$
and triplet $^3S_1$ channels as a function of distance (in fm). The
normalization is such that $u_{0,^1S_0} \to 1-r/\alpha_s$ and $u_{0,^3S_1}
\to 1-r/\alpha_t$ with $\alpha_s = -23.74 {\rm fm}$ and $\alpha_t =
5.42 {\rm fm}$ the singlet and triplet scattering lengths
respectively. The potentials generating these wave functions are the
same $V_s(r)=V_t(r)$.}
\label{fig:u0}
\end{figure}

It is interesting to analyze our results on the light of
Refs.~\cite{Kaplan:1995yg,Mehen:1999qs,Epelbaum:2001fm} where a square
well potential, PDS and sharp momentum cut-off were used respectively
to model the short distance contact interactions arising when all
exchanged particles are integrated out.  Here we are interested in the
dependence on the arbitrary renormalization scale separating the
contact and the extended particle exchange interaction since they are
not independent of each other; by keeping this scale dependence we may
enter the interaction region where, as we will show now, the symmetry
can be visualized. We appeal to the coordinate space version of the
renormalization
group~\cite{PavonValderrama:2004nb,PavonValderrama:2007nu} (for a
momentum space version see Ref.~\cite{Birse:1998dk}), where the
version of the Callan-Zymanzik equation for potential scattering reads
\begin{eqnarray}
R c_0' (R) = c_0 (R) (1-c_0(R)) + M R^2 V(R) \, , 
\label{eq:RG}
\end{eqnarray} 
where $c_0 (R) = R u_0' (R) / u_0 (R) $ is a suitable combination of
the short distance boundary condition and we have chosen for
simplicity to work at zero energy~\footnote{The orthogonality
conditions discussed above correspond to take $c_p (R) \to c_0 (R) $
for $R \to 0$.}. The above equation provides the evolution of the
boundary condition as a function of the distance $R$ (the
renormalization scale) in order to have a fixed scattering amplitude
(see Ref.~\cite{PavonValderrama:2007nu} for a thorough
discussion). Clearly, at long distances $ r \gg 1/m_\pi$ the potential
becomes negligible and the equation is scale invariant, only broken by
the renormalization condition which fixes the value of $c_0$ at some
scale~\footnote{In appendix~\ref{sec:scale} we analyze a case where
the dilatation symmetry of a $1/r^2$ potential {\it must} necessarily
be broken by a renormalization condition.}. In fact, the solution of
the above equation is given in terms of the scattering length
$\alpha_0$ in the infrared, $R \to \infty$ , $c_0 (R) \to \alpha_0
/R $. On the other hand, if the scattering length is large we also
have an intermediate regime with clear scale separation and
\begin{eqnarray} 
c_0 ( R ) = \frac{\alpha_0}{R-\alpha_0} \sim -1  \, , \qquad 1/m_\pi \ll R
\ll \alpha_0 \, ,
\end{eqnarray} 
indicating the onset of scale
invariance~\cite{PavonValderrama:2007nu}. This is in agreement with
the PDS argument of Refs.~\cite{Mehen:1999qs} if the identification
$\mu \sim 1/R$ is done. Eventually the infrared stable fixed point
$c_0 \to 1$ will be achieved. Note, however, that $c_s(R) \sim c_t(R)$
in a much wider range, particularly in the scaling violating region
where the potential acts.  In the more conventional language of wave
functions the situation corresponds to a case where both wave
functions $u_s (r) \sim u_t(r)$ for $r \ll \alpha_s, \alpha_t$. The
situation is illustrated in Fig.~\ref{fig:u0} where the similarity in
the range below 1fm can clearly be seen, and does not differ much from
the solution $u_{0,c} (r)$ entering the superposition principle,
Eq.~(\ref{eq:sup_0}) and corresponding to the limit $\alpha_0 \to \pm
\infty$. Note that the symmetry can be visualized within the range of
the potential only when the scattering length is large because there
exists the scaling regime $1/m_\pi \ll r \ll \alpha_0$, but the long
distance correlations between the two S-wave channels due to the
identity of potentials hold regardless of the unnatural size of the
scattering lengths.

\subsection{Virtual and bound states}

\begin{figure}[tbc]
\includegraphics[height=8cm,width=6.5cm,angle=270]{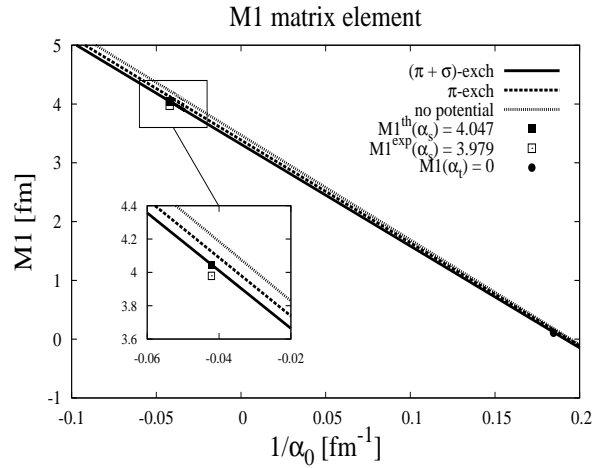}
\caption{ The matrix element ${\cal M} ( \gamma, \alpha) $ for
$\gamma= \gamma_d$ as a function of the inverse scattering length $\alpha_0$.
This function should fulfill the properties ${\cal M} ( \gamma_d,
\alpha_s) = {\cal M}_{M1} $ (the neutron capture M1 matrix element)
and ${\cal M} ( \gamma_d, \alpha_t) = 0 $ (the orthogonality relation
between the deuteron and the zero energy triplet state. The
experimental values are highlighted as points.}
\label{fig:M1}
\end{figure}

It is of course tempting to analyze the kind of features for the
deuteron that may be obtained from this simplified picture where the
tensor force is neglected from the start. The deuteron is determined
by integrating in the Schr\"odinger equation with negative energy
$E=-\gamma_d^2/M$ with $\gamma_d=0.2316 {\rm fm}^{-1}$ the wave number
and imposing the long distance boundary condition,
Eq.~(\ref{eq:bound-long}). We also compute the matter radius
\begin{eqnarray}
r_m^2 = \frac14 \int_0^\infty r^2 u_d (r)^2 \, . 
\end{eqnarray} 
and the ${\cal M}_{M1}$ matrix element
\begin{eqnarray}
A_S {\cal M}_{M1} =  \int_0^\infty dr u_d (r) u_{0,^1S_0}(r)  \, . 
\end{eqnarray} 
which correspond to the dominant magnetic contribution to neutron
capture process $np \to \gamma d$ in the range of thermal neutrons
($\sim {\rm KeV}$) in stars~\footnote{In this normalization the total
cross section is given by
$$ \sigma_M (np \to \gamma d) = \pi \alpha (\mu_p - \mu_n)^2
\sqrt{B/2E} (B/M_N)\gamma {\cal M}_{M1}^2$$ where $E $is the neutron
energy, and $\mu_p$ and $\mu_n$ the proton and neutron magnetic
moments in units of the nuclear magneton, $\mu_N = e /(2 M_p)$. We are
neglecting meson exchange currents in the calculation of ${\cal
M}_{M1}$.}.  For the experimental $\gamma_d = 0.2316 {\rm fm}^{-1}$ we
get $A_S =0.8643 {\rm fm}^{-1/2}$ (exp. $ 0.8846(9) {\rm fm}^{-1/2}$)
and $r_m=1.9138 {\rm fm}$ (exp. $ 1.9754(9) {\rm fm}$ and ${\cal
M}_{M1}= 4.0464 {\rm fm}$) (exp. $ 3.979 {\rm fm}$). As mentioned
above, orthogonality between the bound state and the zero energy state
yields an explicit correlation between the triplet scattering length,
$\alpha_t$ and the deuteron wave number, $\gamma$,
\begin{eqnarray}
\alpha_t = \alpha_0 (\gamma_d ) = 
\frac{\int_0^\infty dr u_\gamma(r) u_{0,s} (r)}
{\int_0^\infty dr u_\gamma (r) u_{0,c} (r) }\Big|_{\gamma=\gamma_d} \, . 
\end{eqnarray} 
Since the two independent zero energy solutions, $u_{0,c} (r)$ and
$u_{0,s} (r)$ depend only on the potential and hence are identical for
the S-wave components of the singlet and triplet channels, this
correlation is a consequence of the Wigner symmetry as well as long as
we take $V_s(r)=V_t(r)$.  Note that taken as a function of the
scattering length, the expression
\begin{eqnarray}
{\cal M} ( \gamma, \alpha_0) = \int_0^\infty dr \, u_\gamma (r) u_0 (r) \, ,
\end{eqnarray} 
yields {\it both} the orthogonality relation as well as ${\cal
M}_{M1}$
\begin{eqnarray} 
{\cal M} ( \gamma_d, \alpha_t) &=& 0 \, , \nonumber \\   
{\cal M} ( \gamma_d, \alpha_s) &=& {\cal M}_{M1} \, .
\end{eqnarray} 
Actually the dependence on the inverse scattering length is a straight
line which we show in Fig.~\ref{fig:M1}. As we see both conditions are
very well fulfilled. Similarly to the previous case, the orthogonality
between finite energy states and the deuteron corresponds to the
magnetic contribution to the photodisintegration of the deuteron. The
result, however does not differ much from the potential-less theory,
and so we will not discuss it any further. For the experimental
$\gamma_d = 0.2316 {\rm fm}^{-1}$ we get $\alpha_t=5.32 {\rm
fm}$. This value improves over the simple formula $\alpha_t = 1/\gamma
= 4.31 {\rm fm}$ obtained from the case without potential, or the
single OPE case where $\alpha_t=4.60 {\rm fm}$. It is worth stressing
that the {\it same} relation above yields the virtual state, a purely
exponentially growing wave function, $u_v (r) \to e^{+\gamma_v r}$, in
the singlet channel, yielding for $\alpha_s = -23.74 {\rm fm}$ -the
value $\gamma_v=0.042 {\rm fm}^{-1}$. In other words, the function
$\alpha_0 (\gamma) $ fulfills $\alpha_0 (\gamma_d) = \alpha_t$ and
{\it simultaneously} $\alpha_0 (\gamma_v) = \alpha_s$.  Numerically we
get
\begin{eqnarray}
\alpha_0 ( -0.042 {\rm fm}^{-1} ) &=& -23.74 {\rm fm}  \, ,\\ 
\alpha_0 ( 0.2265 {\rm fm}^{-1} ) &=&  5.42   {\rm fm}  \, .
\end{eqnarray} 
In the region below 1fm the virtual state $u_v (r) $ and the deuteron
bound state $u_d(r) $ look very much alike the corresponding singlet
and triplet zero energy wave functions respectively (see
Fig.~\ref{fig:u0}).  Thus, $u_{0,^1S_0}(r) \sim u_{v}(r) $ and
$u_{0,^3S_1}(r) \sim u_{d}(r) $ are consequences the closeness of the
poles to the real axis, either in the second or first Riemann sheets
respectively. However, $u_{0,1^S_0}(r) \sim u_{0,3^S_1}(r) $ and $u_{v}(r)
\sim u_{d}(r) $ are further consequences of the identity of the
potentials $V_s(r)=V_t(r)$. 

\section{Symmetry breaking}
\label{sec:symm-break}

\subsection{Symmetry breaking with two counter-terms}
\label{sec:two-ct}

An essential ingredient of the present analysis is the requirement of
orthogonality between different energy states, which ultimately
reflects the self-adjoint character of the Hamiltonian. This implies
that, for the Yukawa like potentials we are dealing with, the only way
to parameterize the unknown information at short distances is by
allowing, besides the regular solution, the irregular one and fixing
the appropriate combination by imposing a value of the scattering
length as an independent renormalization condition. This may appear
too restrictive and in fact it is possible to renormalize using energy
dependent boundary conditions, a procedure essentially equivalent to
imposing more renormalization conditions or counter-terms. Although
there are subtleties on how short distances should be parameterized in
such way that the cut-off may be
removed~\cite{PavonValderrama:2007nu,Entem:2007jg} the procedure in
coordinate space turns out to be rather simple. In the case of two
conditions we would fix the scattering length, $\alpha_0$, and the
effective range $r_0$ independently of the potential.  The coordinate
space procedure~\cite{PavonValderrama:2007nu, Entem:2007jg} consists
of expanding the wave function in powers of the energy
\begin{eqnarray}
u_p (r) = u_0(r) + p^2 u_2 (r) + \dots \, , 
\end{eqnarray} 
where $u_0(r)$ and $u_2(r)$ satisfy the following equations, 
\begin{eqnarray} 
-u_0''(r) + MV (r) u_0(r) &=& 0 \, , \label{eq:u0} \\  
u_0 (r) &\to& 1- r /\alpha_0  \, ,  \nonumber \\ 
-u_2 '' (r) + MV (r) u_2 (r) &=& u_0 (r) \, , \label{eq:u2} \\
u_2 (r) &\to& \left(r^3
-3 \alpha_0 r^2 + 3 \alpha_0 r_0 r \right)/(6 \alpha_0) \, , \nonumber
\end{eqnarray} 
The asymptotic conditions correspond to fix $\alpha_0$ and $r_0$ as
independent parameters (two counter-terms). The matching condition at
the boundary $r=r_c$ becomes energy
dependent~\cite{PavonValderrama:2007nu}
\begin{eqnarray}
\frac{u'_p (r_c)}{u_p(r_c)} = \frac{u'_0 (r_c) + p^2 u'_2 (r_c)+ \dots}{
u_0(r_c)+ p^2 u_2(r_c)+ \dots} \, .
\label{eq:Lp}
\end{eqnarray} 
whence the corresponding phase shift may be deduced by integrating in
Eq.~(\ref{eq:u0}) and Eq.~(\ref{eq:u2}) and integrating out the finite
energy equation. It is worth mentioning that the energy dependent
matching condition, Eq.~(\ref{eq:Lp}), is quite unique since this is
the only representation guaranteeing the existence of the limit $r_c
\to 0 $ for singular potentials~\cite{PavonValderrama:2007nu}. In any
case, if $r_0$ is fixed from the start to their experimental values in
the singlet and triplet channels, the Wigner correlation given by
Eq.~(\ref{eq:r0_univ_st}) and generating the universal curve shown in
Fig.~\ref{fig:r0-a0} would not be predicted and the symmetry between
the $^1S_0$ and the $^3S_1$ channels would be further hidden into the
phase shifts. Note that the breaking of the symmetry with two
counter-terms is a short distance one when the cut-off is eliminated,
$ r_c \to 0$, since at any rate the potential is kept fixed and
$V_s(r) = V_t(r)$ for any non-vanishing distance, $r\ge r_c >0$. Thus,
if we write
\begin{eqnarray}
r_0 = A + \frac{B}{\alpha_0}+ \frac{C}{\alpha_0^2} + r_0^{\rm short} \, , 
\end{eqnarray}
with $r_0^{\rm short} $ the effect of the second counter-term, we would obtain 
\begin{eqnarray}
r_t - r_s \sim r_t^{\rm short}  - r_s^{\rm short} 
+ B \left[\frac1{\alpha_t}-\frac1{\alpha_s}\right] + \dots 
\end{eqnarray}
where small $1/\alpha^2$ terms have been neglected.  This yields
$r_t^{\rm short} - r_s^{\rm short} \sim 0.1 {\rm fm}$.  Thus, while
introducing no counter-term (trivial boundary condition) does not
break the symmetry yielding identical phase shifts,
$\delta_s(k)=\delta_t(k)$, introducing more than one counter-term
(energy dependent boundary condition) breaks the symmetry at the $\sim
10\%$ level. As a consequence, we stick to the case of just one
counter-term (energy independent boundary condition).

\subsection{Symmetry breaking due to the tensor force}
\label{sec:tensor}

Of course, an interesting possibility which should be explored further
is that of keeping the energy independence of the boundary condition
and breaking the symmetry by introducing a long distance component of
the potential, such as e.g. the tensor force, which would include the
coupling of the $^3S_1$ wave treated here to the $^3D_1$ channel.
Actually, this would correspond to take into account, as proposed in
Ref.~\cite{Kaplan:1996rk}, the leading and complete large-$N_c$ NN
potential.  In other words, while Wigner symmetry implies a vanishing
tensor force, leading large-$N_c$ does not necessarily implies the
tensor force to be small. To analyze this potential source of conflict
we consider the $^3S_1$ effective range parameter which incorporates a
D-wave contribution stemming from S-D tensor force mixing and is given
by
\begin{eqnarray} 
r_t &=& 2 \int_0^\infty \left[ \left(1-\frac{r}{\alpha_t} \right)^2 -
u_{0,\alpha} (r)^2 - w_{0,\alpha} (r)^2 \right] dr \, . \nonumber \\ 
\label{eq:r0_triplet} 
\end{eqnarray} 
where the zero energy S-wave function $u_{0,\alpha} (r) \to
u_{0,^3S_1}(r)$ (discussed above) and the D-wave function $w_{0,\alpha}
(r) \to 0 $ when the tensor force is switched off keeping $\alpha_t$
fixed. The corresponding tensor potential would include $\pi$ and
$\rho$ exchange contributions characterized by the $g_{\pi NN}$ and
$f_{\rho NN}$ couplings and diverges as $1/r^3$ at short
distances. This situation resembles a previous OPE
study~\cite{PavonValderrama:2005gu} and a detailed account will be
presented elsewhere~\cite{Calle-LargeN}. There, it will
be shown how the extension of the superposition principle and renormalization
to the
coupled channel case yields  in fact an identical
analytical result as shown in Eq.~(\ref{eq:r0_univ_st}) for the
triplet (un-coupled) channel in the absence of tensor force. We will
just quote here the numerical modification of the correlation relation
coefficients for the triplet channel (the singlet $^1S_0$ is not
modified), Eq.~(\ref{eq:r0_univ_st}).  Numerically, we get for
$f_{\rho NN}=17$ and $g_{\omega NN}=9.86$
\begin{eqnarray}
r_t &=& 2.6199 -
\frac{5.7843}{\alpha_t}+\frac{5.7608}{\alpha_t^2} \, .  
\end{eqnarray} 
which corresponds to a $\sim 10 \%$ breaking due to the tensor force.
As we see, the coefficients in Eq.~(\ref{eq:r0_univ_st_num}) are not
modified much despite the singularity of the tensor force and its
dominance at short distances. Actually, the dependence of the
coefficients on the couplings responsible for the tensor force is
moderate in a wide range. Therefore, while from the large $N_c$
viewpoint a large tensor force is not forbidden, we find the effect in
the S-wave to be numerically small, as implied by Wigner symmetry.

In this regard it should be noted that a
virtue of the renormalization approach is that, since the scattering
lengths are always fixed, such a long distance symmetry breaking term
only influences the region where the potential is resolved, and from
this viewpoint the perturbation will be stable, i.e.  the change will
be small.  Actually, in Ref.~\cite{PavonValderrama:2005gu} a suitable
form of perturbation theory in the tensor force was suggested based on
the known smallness of the mixing angle $\epsilon_1$, which stays
below $2-3^o$, in a wide energy range and is indeed smaller than the
$\delta_\beta$ phase. It would be interesting to work out the
consequences of such an approach when also $\rho$ exchange is
incorporated.

\subsection{Symmetry  breaking in non-central waves}
\label{sec:non-central}

With the previous appealing interpretation of the Wigner symmetry as a
long distance one for the S-waves, we analyze what are the
consequences for the phase shifts corresponding to partial waves at
angular momentum larger than zero, $L> 0$. Unlike the S-waves we
expect the dependence on the short distance behavior to be suppressed
due to the centrifugal barrier, and the symmetry should become more
evident. Note also that while a dissimilarity between phase shifts
connected by the symmetry does not necessarily imply long distance
symmetry breaking, an identity between phase shifts is a clear hint of
the symmetry.

In the two-nucleon system the Wigner symmetry implies the following
relations for spin-isospin components of the antisymmetric sextet,
${\bf 6}_A$, and the symmetric decuplet, ${\bf 10}_S$, respectively
(see Appendix~\ref{eq:su4}) thus we should have 
\begin{eqnarray}
\delta_{LJ}^{01} &=& \delta_{LJ}^{10} = \delta_L \, , \qquad {\rm
even}-L \\ \delta_{LJ}^{00} &=& \delta_{LJ}^{11} = \delta_L \, ,
\qquad {\rm odd}\, -L
\end{eqnarray} 
For P-waves, for instance, we have the spin singlet state $^1P_1$ and
the spin triplets $^3P_0$,$^3P_1$ and $^3P_2$ which according to the
symmetry should be degenerate as they belong to the ${\bf 10}_S$
supermultiplet. Inspection of the Nijmegen
analysis~\cite{Stoks:1993tb} reveals that $^1P_1$ is very similar to
$^3P_1$ at all energies, $|\delta_{^1P_1} - \delta_{^3P_1} | \sim 1^0
$, but very different from the $^3P_0$ and $^3P_2$ phases. For
D-waves, associated to a ${\bf 6}_A$ supermultiplet, we have a
similarity between $^1D_2$ and $^3D_3$ phases $|\delta_{^1D_2} -
\delta_{^3D_3} | \sim 1^0 $ but, again, clear differences between the
$^3D_1$ and $^3D_2$ ones. Clearly, the symmetry is broken in higher
partial waves. In what follows we want to determine whether
our interpretation of a long distance symmetry
which worked so successfully for S-waves above (see
Sect.~\ref{sec:numeric}) holds also for non-central phases.

As it is well-known the spin-orbit interaction lifts
the independence on the total angular momentum, via the operator $\vec
L \cdot \vec S$. Moreover, the tensor coupling operator, $S_{12}$, 
mixes states with different orbital angular momentum. We proceed in
first order perturbation theory, by using the Wigner symmetric
distorted waves as the unperturbed states. In
appendix~\ref{sec:splitting} we show this procedure explicitly. To
first order in spin-orbit and tensor force perturbation the following
sum rule for the center of the $S=1$ multiplet, denoted by 
$\delta_{L}^{10}$ and $\delta_{L}^{11}$, and the $S=0$ states,
denoted as $\delta_{L}^{01}$ and $\delta_{L}^{00}$, holds,
\begin{eqnarray}
\delta_{L}^{10} &\equiv& \frac{\sum_{J=L-1}^{L+1} (2J+1)
\delta_{LJ}^{10}}{(2L+1)3} = \delta_{LL}^{01} \equiv \delta_{L}^{01}
\, , \nonumber \\ 
\delta_{LL}^{11} &\equiv & \frac{\sum_{J=L-1}^{L+1} (2J+1)
\delta_{LJ}^{11}}{(2L+1)3}  = \delta_{LL}^{00} \equiv \delta_{L}^{00}\, ,
\label{eq:ps-lande}
\end{eqnarray}
In terms of these mean phases, Wigner symmetry is formulated for
non-central waves as
\begin{eqnarray}\
\delta_{^1P_1} &=& \frac{1}{9} \left(  \delta_{^3P_0} + 3
\delta_{^3P_1} + 5 \delta_{^3P_2} \right) \\ 
\delta_{^1D_2} &=& \frac{1}{15} \left( 3 \delta_{^3D_1} + 5
\delta_{^3D_2} + 7 \delta_{^3D_3} \right) \\
\delta_{^1F_3} &=& \frac{1}{21} \left( 5 \delta_{^3F_2} + 7
\delta_{^3F_3} + 9 \delta_{^3F_4} \right) \\
\delta_{^1G_4} &=& \frac{1}{27} \left( 7 \delta_{^3G_3} + 9
\delta_{^3G_4} + 11 \delta_{^3G_5} \right) \, . 
\label{eq:sr-wigner}
\end{eqnarray} 
\begin{figure*}[tbc]
\includegraphics[height=8cm,width=6.5cm,angle=270]{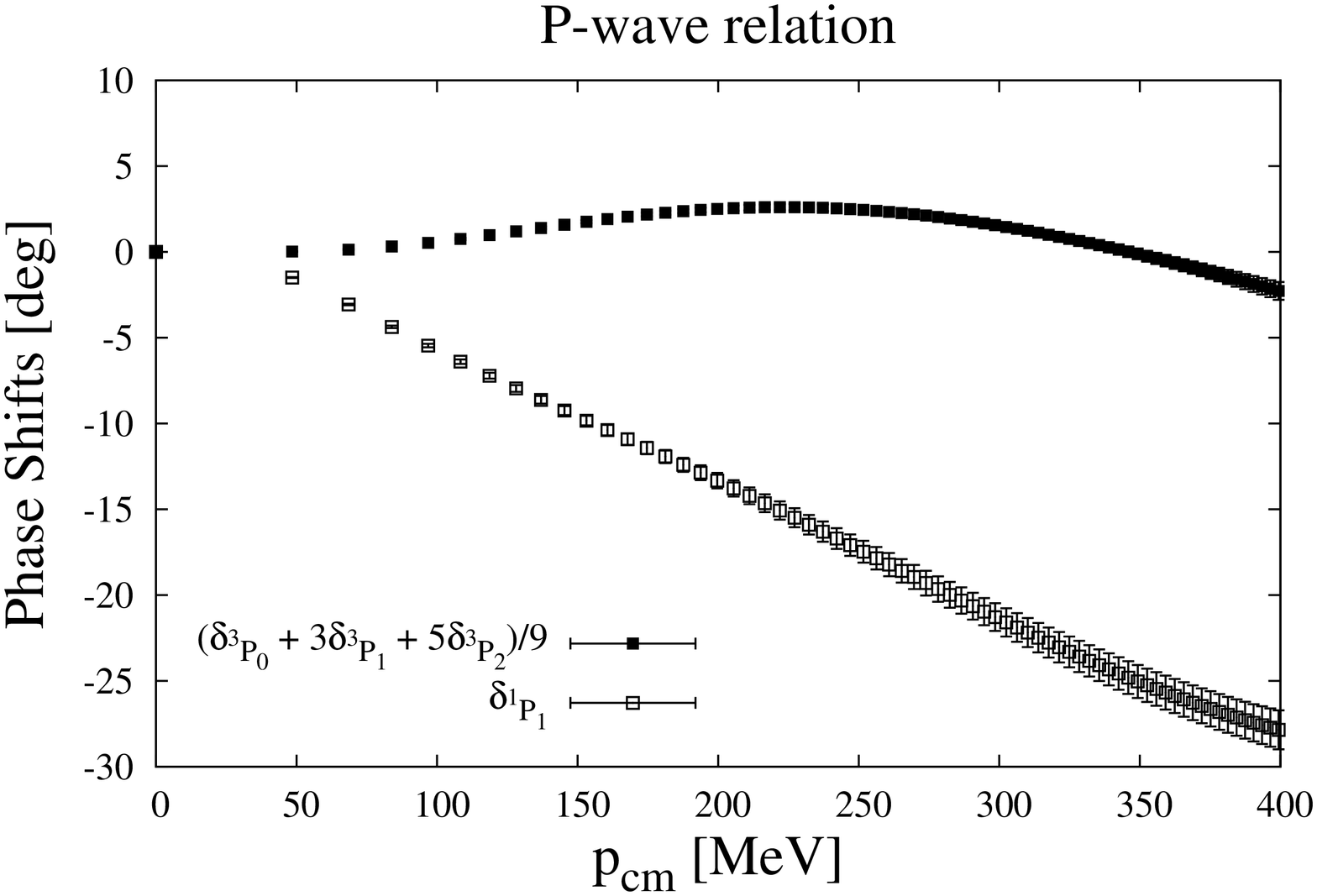}
\includegraphics[height=8cm,width=6.5cm,angle=270]{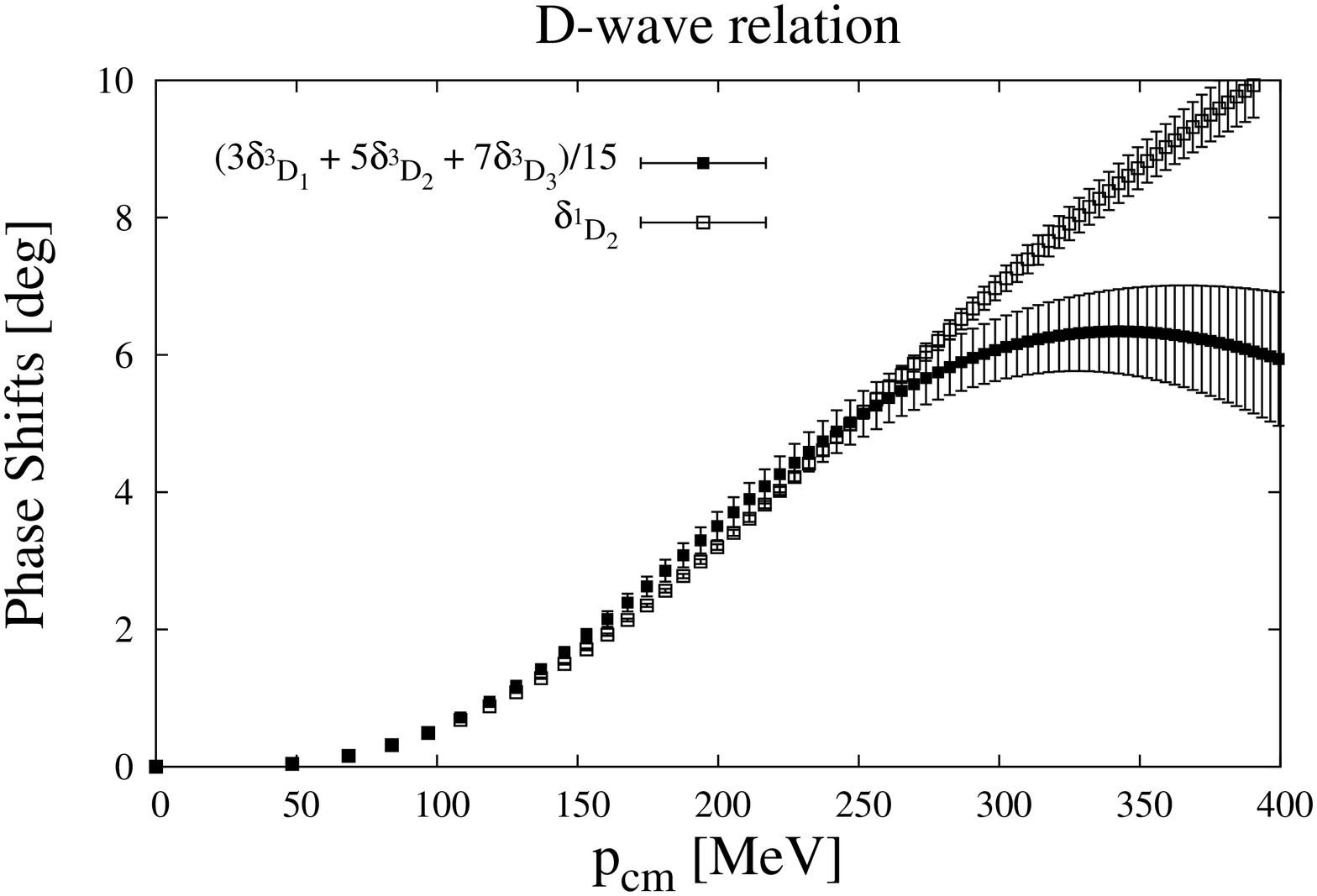} \\ 
\includegraphics[height=8cm,width=6.5cm,angle=270]{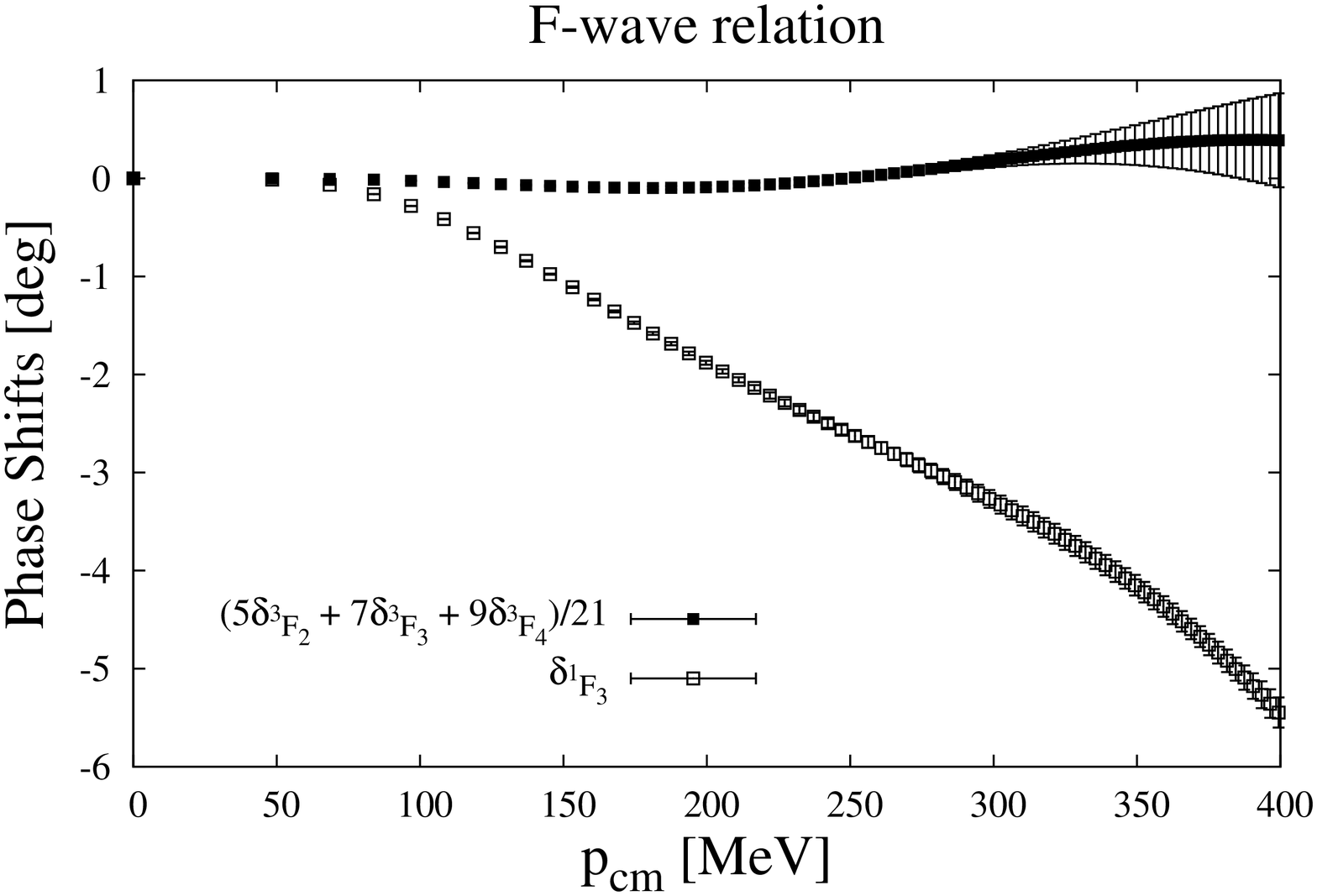}
\includegraphics[height=8cm,width=6.5cm,angle=270]{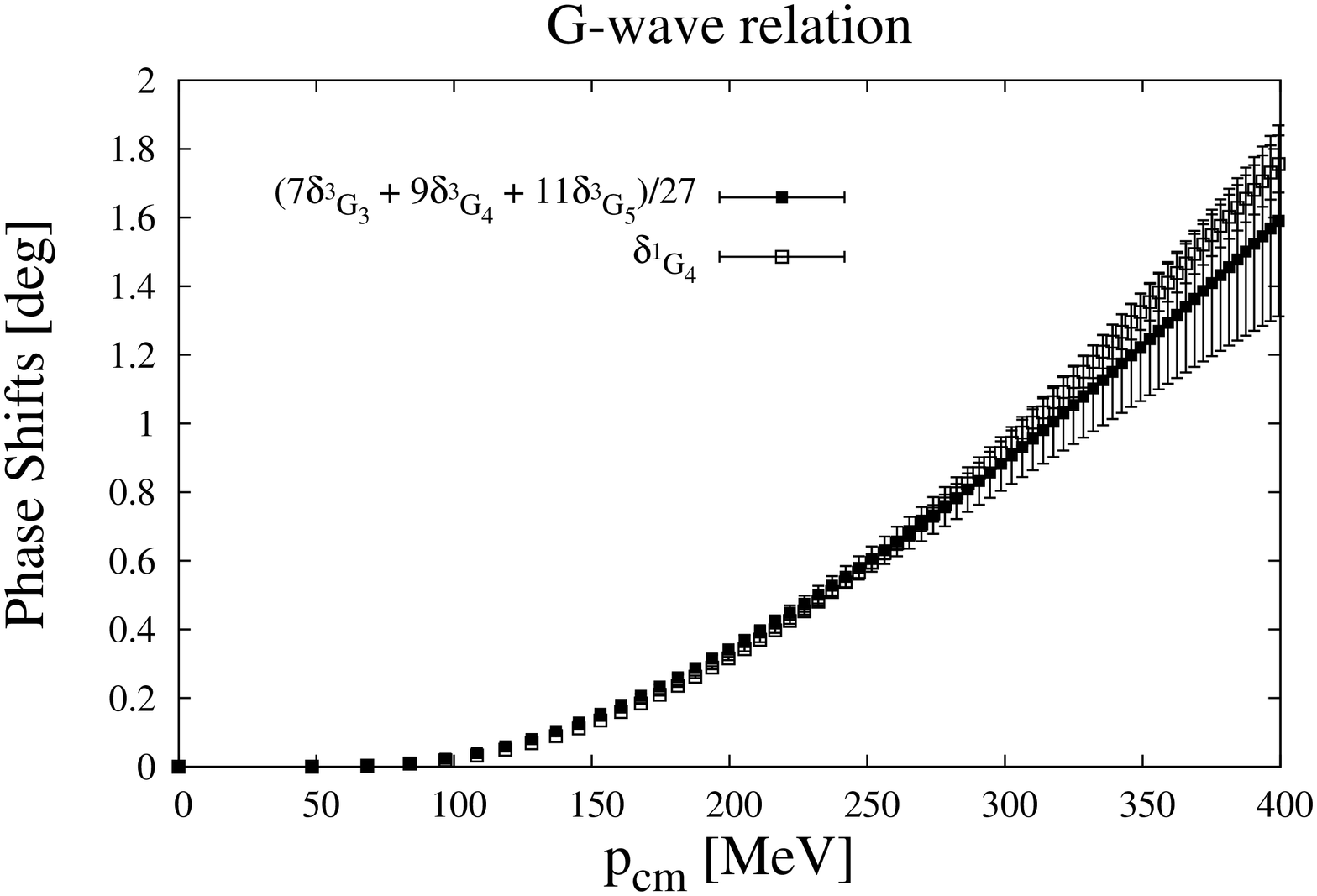}
\caption{Average values of the phase shifts~\cite{Stoks:1993tb} (in
degrees) as a function of the CM momentum (in MeV) based on first
order spin-orbit coupling.  (Upper left panel) P-waves. (Upper right
panel) D-waves. (Lower left panel) F-waves. (Lower right panel)
G-waves. According to the Wigner symmetry
$\delta_{^1L}=\delta_{^3L}$. Serber symmetry implies $\delta_{^3L}=0$
for odd-L. One sees that L-even waves satisfy Wigner symmetry while L-odd
waves satisfy Serber symmetry.}
\label{fig:higher}
\end{figure*}
These sum rules are true as long as the short distance breaking can be
considered small, and for this reason we have not written down the sum
rule for S-waves. Further, they hold also when the tensor force is
added. In Fig.~\ref{fig:higher} we show the l.h.s. and the r.h.s. of
P-, D-, F- and G-waves. As we see the D-waves fulfill this relation
rather accurately up to $p \sim 250 {\rm MeV}$ and the G-waves up to
$p \sim 400 {\rm MeV}$ while the P- and F-waves fail
completely. Actually, at threshold, $\delta_L \to - \alpha_L
p^{2L+1}$, and using the low energy parameters of the NijmII and
Reid93 potentials~\cite{Stoks:1993tb} determined in
Ref.~\cite{PavonValderrama:2005ku} we get
\begin{eqnarray}
\alpha_{^1P_1} &=& \frac{1}{9} \left(  \alpha_{^3P_0} + 3
\alpha_{^3P_1} + 5 \alpha_{^3P_2} \right) \, ,   \nonumber \\ 
(-2.46 {\rm fm}^3) && (0.08 {\rm fm}^3) \, ,   \nonumber \\
\alpha_{^1D_2} &=& \frac{1}{15} \left( 3 \alpha_{^3D_1} + 5
\alpha_{^3D_2} + 7 \alpha_{^3D_3} \right) \, ,   \nonumber \\ 
(-1.38 {\rm fm}^5) && (-1.23 {\rm fm}^3)  \, ,  
\end{eqnarray} 
where the numerical values are displayed below the sum rules. On
the light of the previous discussions for the S-waves one reason for
the discrepancy should be looked in a short distance breaking of the
symmetry for the D-waves. Actually, the fact that D-waves violate the
sum rule at $p \sim 250 {\rm MeV}$ while the G-waves show no violation
up to $p \sim 400 {\rm MeV}$ agrees with our interpretation in the
S-waves that the Wigner symmetry be a long distance one, since higher
partial waves are less sensitive to short distance effects.  The case
of P-waves is different since the $^1P$-potential and the
$^3P$-potentials are very different. This pattern of symmetry breaking
agrees with the findings of Ref.~\cite{Kaplan:1996rk} based on the
large $N_c$ expansion where the central potential preserves the
symmetry in $L$-even partial waves while it breaks the symmetry in the
$L$-odd partial waves, since at leading order and neglecting the
tensor force
\begin{eqnarray} 
V(r)= V_C (r) + \sigma \tau W_S (r) + {\cal O} (1 /N_c) \, ,  
\end{eqnarray} 
so that for the lower L-channels we have 
\begin{eqnarray}
V_{^1S} = V_{^3S} \, ,  &=&   
V_C (r) - 3 W_S (r) + {\cal O} (1 /N_c) \, ,  \nonumber  \\  
V_{^1P} &=&   V_C (r) + 9 W_S (r) + {\cal O} (1 /N_c) \, ,  \nonumber  \\  
V_{^3P} &=&   V_C (r) +  W_S (r) + {\cal O} (1 /N_c) \, ,   \nonumber  \\  
V_{^1D} = V_{^3D} &=&   V_C (r) - 3 W_S (r) + {\cal O} (1 /N_c) \, ,  \nonumber  \\  
\end{eqnarray} 
so as we see $V_{^3P} \neq V_{^1P}$, and thus it is obvious that
$\delta_{^3P} \neq \delta_{^1P}$. One might check this further by
proceeding as follows. In the case of odd waves such as the P-waves
the proper comparison might be taking the $^3P$-potential and
renormalizing with the $^3P$-mean scattering length, $\alpha_{^3P}=
0.08 {\rm fm}^3$ and compare to the $^3P$-mean phase shift. 

We note that the initial claim of Ref.~\cite{Kaplan:1995yg} on the
validity of the Wigner symmetry based on the large $N_c$ expansion was
restricted to purely center potentials, which do not faithfully
distinguish the two irreducible representations, ${\bf 10}_S$ and
${\bf 6}_A$, of the $SU(4)$ group for the NN system. Later on, the
issue was qualified by a more complete study carried out in
Ref.~\cite{Kaplan:1996rk} which in fact could not justify the Wigner
symmetry in odd-L partial waves, even when the tensor force was
neglected. Although this appeared as a puzzling result, it is amazing
to note that our calculations clearly show that the pattern of
$SU(4)$-symmetry breaking supports a weak violation in even-L partial
waves and a strong violation in the odd-L partial waves, {\it exactly
} as the large $N_c$ expansion suggests.

\subsection{Serber symmetry}

On the other hand, from the odd-waves we see from
Fig.~\ref{fig:higher} that the mean triplet phase is close to null,
thus one might attribute this feature to an accidental symmetry where
the odd-waves potentials are likewise negligible. In the large $N_c$
limit this means $V_C + W_S \gg V_C + 9 W_S$, a fact which is well
verified. For instance at short distances the Yukawa OBE potentials have
Coulomb like behavior $V \to C/(4 \pi r)$ with the dimensionless
combinations
\begin{eqnarray}
C_{V_C+W_S}&=&-g_{\sigma NN}^2 +g_{\omega NN}^2 + 
 \frac{ f_{\rho NN}^2 m_\rho^2 }{6 M_N^2}
 \nonumber \\
C_{V_C+9W_S}&=&-g_{\sigma NN}^2 +g_{\omega NN}^2 + 
 \frac{ 3 f_{\rho NN}^2 m_\rho^2 }{2 M_N^2}
 \nonumber \\
\end{eqnarray}
where the small OPE contribution has been dropped.  Numerically we get
$C_{V_C+W_S} \sim 10 $ and $C_{V_C+9W_S} \sim 300 $ for reasonable
choice of couplings. Although this approximate vanishing of triplet
odd-wave potentials {\it is not} a consequence of large $N_c$
it is nevertheless reminiscent of the old and
well-known Serber force,
\begin{eqnarray}
V_{\rm Serber} (r) &=& \frac12 \left( 1+ P_M \right) \frac12 \left( 1-
 P_\sigma \right) V_s(r) \nonumber \\ &+&\frac12 \left( 1- P_M \right)
 \frac12 \left( 1+ P_\sigma \right) V_t(r) \, , 
\end{eqnarray} 
with $P_M$ the Majorana coordinate exchange operator. Due to the Pauli
principle $P_M P_\sigma P_\tau = -1 $ with $P_\tau= (1+\tau)/2$ and
$P_\sigma= (1+\sigma)/2$ the isospin and spin exchange yields
vanishing potentials for spin-triplet and isospin-triplet channels,
and generating a scattering amplitude which is even in the CM
scattering angle, a property which is approximately well fulfilled
experimentally for pp-scattering. We call this property Serber
symmetry for definiteness. After introducing spin-orbit coupling we
would get the sum rules to first order
\begin{eqnarray}
\delta_{^3P} &\equiv& \frac{1}{9} \left(  \delta_{^3P_0} + 3 
\delta_{^3P_1} + 5 \delta_{^3P_2} \right) =0 \, , \\ 
\delta_{^3F} &\equiv& \frac{1}{21} \left( 5 \delta_{^3F_2} + 7
\delta_{^3F_3} + 9 \delta_{^3F_4} \right) =0 \, , 
\label{eq:sr-serber}
\end{eqnarray} 
which is well fulfilled by the phase shifts~\cite{Stoks:1993tb} as
shown in Fig.~\ref{fig:higher} where $\delta_{^3P} \ll \delta_{^1P} $
and $\delta_{^3F} \ll \delta_{^1F} $. In the large $N_c$ limit we may
comply both with Wigner symmetry in L-even waves and Serber symmetry
in L-odd waves when $W_S (r)=-V_C (r)$, whence generally $V (r) = V_C
(r) (1- \sigma \tau)$. Even if we neglect the small OPE effects, this
will clearly not be exactly fulfilled unless one would require $m_\rho
= m_\omega = m_\sigma$. Although there are schemes where such an
identity between scalar and vector meson masses are explicitly
verified~\cite{Weinberg:1990xn,Svec:1996xp,Megias:2004uj}, at present,
it is unclear whether the Serber symmetry which we observe in the NN
system for spin-triplet and odd-L phase-shifts could be formulated as
a symmetry from the underlying QCD Lagrangean.

Our findings suggest that a pure large $N_c$ in the absence of tensor
force not only is compatible with the standard Wigner symmetry in the
case of the dominant S-waves and higher L-even channels, but it might
also be a competitive alternative for the L-odd waves where the usual
Wigner symmetry is broken and Serber symmetry holds instead. Of course
it would be interesting to pursue the more complete situation
including the tensor force from the start, a case which will be
presented elsewhere~\cite{Calle-LargeN}.

\subsection{NN Level density in the continuum}

Our results have some impact for hot nuclear matter at low densities.
In the continuum, we may think of putting the two nucleon system in a
box and evaluate the corresponding level density when the infinite
volume limit is taken. This is a standard problem in statistical
mechanics which appears, e.g. in the calculation of the second virial
coefficient contribution to the equation of state of a dilute quantum
gas~\cite{Dashen:1969ep} (see
Refs.~\cite{Horowitz:2005nd,Mallik:2008zt} for recent applications to
hot nuclear matter). The result is expressed as
\begin{eqnarray}
\rho(E) = \frac1{2\pi i} {\rm Tr} \left[S(E)^\dagger \frac{d S(E)}{dE}
\right]= \frac{1}{\pi} \frac{d\Delta_{NN} (E)}{dE} \, , 
\end{eqnarray} 
where $S(E)$ is the S matrix in all coupled channels and the 
total phase $\Delta$ is defined by 
\begin{eqnarray}
\Delta_{NN} (E) = \sum_{S,T,J} (2 J+1) (2T+1) \delta_{LJ}^{ST} (E) \, . 
\end{eqnarray} 
In the case of coupled channels one should consider the corresponding
eigenphases~\footnote{In the special case of NN scattering one can
also use the nuclear bar phase shifts due to the identity $ \bar
\delta_{^3 (J-1)_J} + \bar \delta_{^3 (J+1)_J} = \delta_{^3 (J-1)_J} +
\delta_{^3(J+1)_J} $. The concern spelled out in
Ref.~\cite{Horowitz:2005nd} that neglecting the mixing was an
approximation is unjustified.}.  
Defining the mean phase as 
\begin{eqnarray}
\delta_{L}^{ST} (E) \equiv \frac{\sum_{J=L-S}^{L+S} (2J+1)
\delta_{LJ}^{ST} (E)
}{(2S+1)(2L+1)} \, , 
\label{eq:ps-mean} 
\end{eqnarray} 
corresponding to the phase-shift analog of the center of gravity of
the supermultiplet (see also Eq.~(\ref{eq:ps-lande}) we get
\begin{eqnarray}
\Delta_{NN} (E) = \sum_{S,T,J} (2S+1) (2 L+1) (2T+1) \delta_{L}^{ST} (E) \, .
\end{eqnarray} 
Thus, using the above relations, Eq.~(\ref{eq:sr-wigner}) for L-even
waves and Eq.~(\ref{eq:sr-serber}) for L-odd waves, featuring Wigner
and Serber symmetries respectively we would get that mixed triplet
channel contributions may be either eliminated in terms of singlet
ones for even-L or do not contribute for odd-L,
\begin{eqnarray}
\Delta_{NN} (E) = 3 \left(\delta_{^1S_0} + \delta_{^3S_1} \right) + 3
\delta_{^1P_1} + 30 \delta_{^1D_2} + \dots 
\end{eqnarray} 
For the neutron case we have 
\begin{eqnarray}
\Delta_{nn} (E) = \delta_{^1S_0} + 5 \delta_{^1D_2} + 9 \delta_{^1G_4}
+ \dots
\end{eqnarray} 
i.e., odd-L waves do not contribute. The lack of a P-wave contribution
scaling as $\sim -\alpha_P p^3$ is compatible with the minimum 
observed in Ref.~\cite{Horowitz:2005nd} for $\Delta_{nn}$ in the
subthreshold region $E_{\rm LAB} < 50 {\rm MeV}$.

\section{Conclusions}
\label{sec:concl}

At low energies NN interactions are dominated by two S-waves in
different channels where spin-isospin $(S,T)$ are interchanged, $(1,0)
\leftrightarrow (0,1)$. Wigner $SU(4)$ symmetry implies that the
potentials in the $^1S_0$ and $^3S_1$ channels coincide and the tensor
force vanishes, while the corresponding phase shifts from Partial Wave
Analyses are actually very different at all energies and show no
evident trace of the identity of the potential, besides the
qualitative fact that a weakly bound deuteron $^3S_1$ state and an
almost bound virtual $^1S_0$ take place. Given the fact that the
nuclear force at short distances is fairly unknown, the validity of
the symmetry to all distances would be at least questionable and could
hardly be tested quantitatively. On the other hand, our lack of
knowledge of the short distance physics should not be crucial at low
energies, where the phase shifts are indeed quite
dissimilar. Therefore, we propose to regard $SU(4)$ as a long distance
symmetry which might be strongly broken at short distances and weakly
broken at large distances. Using renormalization ideas where the
desirable short distance insensitivity is manifestly fulfilled we have
shown how the standard Wigner correlation between potentials indeed
predicts one phase shift from the other in a non-trivial and
successful way. Remarkably, using a large $N_c$ motivated One Boson
Exchange potential we have proven that if one channel is described
successfully the other channel is unavoidably well reproduced within
uncertainties which might be compatible with the disregard of the
tensor force and the $1/N_c^2$ corrections to the potential. This long
distance correlation holds also for the virtual singlet state and the
deuteron bound state.  Actually, the effects of symmetry breaking at
long and short distances have been analyzed and the extension to
higher partial waves has also been discussed, where a relation for
phase shifts has been deduced.  Our calculations provide a
justification on the use of Wigner symmetry in light nuclei solely on
the basis of the NN-interaction and suggest that a specific
interpretation of the Wigner symmetry as a long distance one in
conjunction with renormalization theory extends beyond the scaling
region to a much wider range than assumed hitherto. It would be
interesting to see how these ideas could be further exploited beyond
the simple two nucleon system.  However, a key question has always
been what is the origin of the accidental Wigner symmetry from the
underlying fundamental QCD Lagrangean and, moreover, under what
conditions this is expected to be a useful symmetry. We find the large
$N_c$ expansion in the absence of tensor force besides being
compatible with the standard Wigner symmetry in the case of the low
energy dominant S-waves and subdominant higher $L$-even partial waves
it may also become a competitive alternative for the other $L$-odd
partial waves where the usual Wigner symmetry is manifestly
broken. These conclusions are remarkable, for they suggest that a
unforeseen handle on the nature, applicability and interpretation of a
widely used approximate nuclear symmetry may be based on a QCD
distinct pattern such as the large $N_c$ limit. Obviously, it would be
very interesting to pursue further the study of the complete large
$N_c$ potential with inclusion of the tensor force to verify this
issue in more detail~\cite{Calle-LargeN}. In our view this would
definitely provide useful insights into QCD inspired approximation
schemes in nuclear physics.

\begin{acknowledgements} 
We gratefully acknowledge Manuel Pav\'on Valderrama and Daniel
Phillips for critical remarks on the ms.  A. C. C. thanks Robin
C\^ot\'e for his hospitality in Storrs where part of this work has
been done.  This work has been partially supported by the Spanish DGI
and FEDER funds with grant FIS2005-00810, Junta de Andaluc{\'\i}a
grant FQM225-05, and EU Integrated Infrastructure Initiative Hadron
Physics Project contract RII3-CT-2004-506078.
\end{acknowledgements} 

\appendix

\section{Wigner symmetry for NN}
\label{eq:su4}

Wigner $SU(4)$ spin-isospin symmetry consists of the following
15-generators~~\cite{1941RPPh....8..274W,Wilkinson:1969,
1999RPPh...62.1661V}
\begin{eqnarray}
T^a &=& \frac12 \sum_A\tau_A^a  \, ,  \\ 
S^i &=& \frac12 \sum_A \sigma_A^i \, ,  \\ 
G^{ia} &=& \frac12 \sum_A \sigma_A^i \tau_A^a \, , 
\end{eqnarray} 
where $\tau_A^a$ and $\sigma_A^i$ are isospin and spin Pauli matrices
for nucleon $A$ respectively, and $T^a$ is the total isospin, $S^i$
the total spin and $G^{ia}$ the Gamow-Teller transition operator.  The
quadratic Casimir operator reads
\begin{eqnarray}
C_{SU(4)} &=& T^a T_a + S^i S_i + G^{ia} G_{ia} \, , 
\end{eqnarray} 
and a complete set of commuting operators can be taken to be
$C_{SU(4)}$, $T_3$ and $S_z, G_{z3}$.  The fundamental
representation has $C_{SU(4)}=4$ and corresponds to a single nucleon
state with a quartet of states $p\uparrow$, $p\downarrow$,
$n\uparrow$, $n\downarrow$, with total spin $S=1/2$ and isospin
$T=1/2$ represented ${\bf 4}= (S,T)=( 1/2,1/2)$. For two nucleon
states with good spin $S$ and good isospin $T$ Pauli principle
requires $(-)^{S+T+L}=-1$ with $L$ the angular momentum, thus
\begin{eqnarray}
C_{SU(4)}^{ST} = \frac12\left(\sigma + \tau + \sigma \tau\right) +
\frac{15}2 \, , 
\end{eqnarray} 
where $\tau= \tau_1 \cdot \tau_2 = 2 T(T+1)-3 $ and $\sigma= \sigma_1
\cdot \sigma_2 = 2 S (S+1)-3 $ and the corresponding wave function is
of the form
\begin{eqnarray}
\Psi ( \vec x ) = \frac{u_{L}^{LS} (r)}{r} Y_{L M_L}(\hat x) \chi^{S M_S}
\chi^{T M_T} \, . 
\end{eqnarray} 
One has two supermultiplets, which Casimir values are  
\begin{eqnarray}
C_{SU(4)}^{00} &=& C_{SU(4)}^{11} = 9  \, , \\ 
C_{SU(4)}^{01} &=& C_{SU(4)}^{10} = 5  \, , 
\end{eqnarray} 
corresponding to an antisymmetric sextet ${\bf 6}_A=(0,1) \oplus
(1,0)$ when $L=$ even and a symmetric decuplet ${\bf 10}_S=(0,0)
\oplus (1,1)$ when $L=$odd. The radial wave functions fulfill
$u_L^{01}(r)=u_L^{10} (r)$ and $u_L^{00}(r)=u_L^{11} (r)$
respectively.  This means that we have the following supermultiplets
\begin{eqnarray}
(^1S_0, ^3S_1 ) \,, \,  (^1P_1, ^3P_{0,1,2})  
\,, \,  (^1D_2, ^3D_{1,2,3}) \dots  
\end{eqnarray} 
When applied to the NN potential, the requirement of Wigner symmetry
for {\it all} states, implies
\begin{eqnarray}
V_T &=&W_T=V_{LS}=W_{LS}=0 \, , \nonumber \\ W_S &=& V_S=W_C \, ,  
\end{eqnarray} 
so that the potential may be written as 
\begin{eqnarray} 
V = V_C + (2C_{SU(4)}^{ST}-15) W_S  \, .  
\end{eqnarray} 
Note that the particular choice $W_S=0$ corresponds to a spin-isospin
independent potential, but in this case no distinction between the
${\bf 6}_A$ and ${\bf 10}_S$ supermultiplets arises. As it is
well-known the spin-orbit interaction lifts the total angular momentum
independence. The Wigner symmetry does not distinguish between
different total angular momentum values, so admitting that the
potentials are different we may define a common potential
\begin{eqnarray}
V_{LST} (r) \equiv \frac{\sum_{J=L-S}^{L+S} (2J+1)
V_{JST} (r)
}{(2S+1)(2L+1)} \, , 
\end{eqnarray} 
where similarly to the perturbation theory for energy levels where the
center of a multiplet of states is predicted, the appropriate
statistical weights related to the angular momentum have been
used. The previous expression makes sense if the symmetry is broken
linearly by spin-orbit coupling. In terms of these mean potentials
the symmetry would be
\begin{eqnarray}
V_{^1L} (r) =  V_{^3L} (r) \, ,  
\end{eqnarray} 
or equivalently 
\begin{eqnarray}
V_{^1J_J} (r) =  \frac{\sum_{J=L-1}^{L+1} (2J+1)
V_{^3L_J} (r)}{3(2L+1)} \, .
\end{eqnarray} 
As mentioned in the paper, if the symmetry is taken literally at {\it
all} distances we should have $\delta_{^1L}=\delta_{^3L}$.

\section{Long distance Perturbation theory}
\label{sec:pert}

We illustrate here a situation where the potential may be treated in
long distance perturbation theory and renormalized (for a somewhat
similar approach for finite cut-offs see
e.g. Ref.~\cite{Cohen:1998bv}). Unlike the standard perturbative
approach, which usually does not hold in the presence of bound states,
this expansion can deal with weakly bound states, provided this is the
only one. This is in fact the case for the OPE potential for the
parameters we use, applied to the deuteron state, for which we show
the procedure here to first order.  To analyze this situation we vary
the potential $V \to V+ \Delta V$
\begin{eqnarray}
- \Delta u_k (r) '' &+& M\Delta V(r) u_k(r) \nonumber \\ 
&+& M V(r) \Delta u_k(r)= k^2
  \Delta u_k (r) \, , 
\label{eq:duuu}
\end{eqnarray} 
we use the previous wave functions $u_k(r)$ as the zeroth order
approximation, corresponding to take $V(r)=0$ and solve for the first
order correction $\Delta u_k(r)$ the equation which asymptotic wave
function corresponds to take the phase shift $\delta + \Delta
\delta$. Multiplying Eq.~\eqref{eq:Scrod-p} by $\Delta u_k(r)$ and
Eq.~\eqref{eq:duuu} by $u_k(r)$, subtracting both equations and
integrating from $r_c$ to $\infty$ we get 
\begin{eqnarray}
\left[-u_k' \Delta u_k + u_k \Delta u_k \right]\Big|_{r_c}^\infty =
\int_{r_c}^\infty dr \, \Delta U(r) u_k (r)^2 \, .
\end{eqnarray} 
The lower limit term may be related to the variation of the boundary
condition, whereas the upper limit term is related to the change in
the phase shift, $\Delta \delta$. In order to eliminate the cut-off we
subtract the zero energy limit, $ k\to 0$, and using the energy
independence of the boundary condition we get some cancellation since 
\begin{eqnarray}
\Delta \left( \frac{u_k'(r_c)}{u_k(r_c)} - \frac{u_0'(r_c)}{u_0(r_c)}
\right) =0 \, .
\label{eq:orth-pert}
\end{eqnarray} 
Finally, the  result may be re-written as follows  
\begin{eqnarray}
\Delta \left( k \cot \delta \right) &=& -\Delta \left( \frac1{\alpha_0}
\right)  \\ &+& \int_{r_c}^\infty \Delta U(r) \left[ u_k
(r)^2 - u_0 (r)^2 \right] dr \, . \nonumber 
\label{eq:deltacot}
\end{eqnarray} 
If we fix the scattering length independently on the potential we have
$\Delta \alpha_0=0$, thus eliminating the first term of the r.h.s. and
after taking the limit $r_c \to 0$ the result for the total (and
renormalized) phase shift to first order in the potential reads
\begin{eqnarray}
k \cot \delta_0 (k) &=&-\frac1{\alpha_0} + \int_{0}^\infty dr \, M V(r)
 \\ &\times& \left( \left[ \cos(kr ) -\frac{\sin(k
r)}{\alpha_0 k} \right]^2 - \left[ 1-\frac{r}{\alpha_0} \right]^2
\right) + \dots \nonumber 
\end{eqnarray}
The renormalized effective range is {\it entirely predicted} from the
potential at all distances
\begin{eqnarray}
r_0 = 4 \int_{0}^\infty dr r^2 M V(r) \left( 1-
\frac{r}{\alpha_0}\right)^2 + \dots
\label{eq:r0_pert} 
\end{eqnarray}
Note the extra power suppression at the origin when $\alpha_0$ is
fixed independently on the potential, indicating short distances
become {\it less} important. The bound state can be obtained in a
similar manner by replacing $u_k(r) \to u_\gamma (r)$, assuming that
the binding energy is independent on the potential, $\Delta \gamma =0
$, and using orthogonality Eq.~(\ref{eq:orth-pert}) to the zero energy
state
\begin{eqnarray}
\frac1{\alpha_0} = \gamma + \int_{0}^\infty  M V(r) \left[ u_\gamma
(r)^2 - u_0 (r)^2 \right] dr \, . 
\end{eqnarray} 
This equation is implicit in both $\alpha_0$ and $\gamma$, but we can
make it perturbative explicitly, using that to first order $\alpha_0
\sim 1/\gamma$ in the zero energy wave function $u_0 (r) \sim 1 -
\gamma r $, yielding
\begin{eqnarray}
\frac1{\alpha_0} = \gamma + \int_{0}^\infty M V(r) \left[e^{-2 \gamma
r}- (1-\gamma r)^2 \right] dr
\label{eq:alpha0-gamma}
\end{eqnarray}   

\section{Scale invariance and renormalization}
\label{sec:scale}

We have suggested that Wigner symmetry be a long distance one. From a
renormalization group (RG) viewpoint this has a simple interpretation (for
a discussion in coordinate space see
e.g. Ref.~\cite{PavonValderrama:2004nb,PavonValderrama:2007nu}).  It
means finding a solution to the RG equations which break the symmetry
of the equations. A very simple case which illustrates this issue is
provided by the problem
\begin{eqnarray}
-u'' (r) + \frac{g}{r^2} u(r) = k^2 u(r) \, .
\end{eqnarray} 
At zero energy, $k=0$, the solution is invariant under the scaling
transformation $r \to \lambda r $. This property holds also at short
distances, where the energy term on the r.h.s. can be neglected. If we
use the RG equation, Eq.~(\ref{eq:RG}), for this particular case at
short distances 
\begin{eqnarray}
R c_0' (R) = c_0 (R) (1-c_0(R)) + g \, . 
\end{eqnarray} 
The scale symmetry becomes now evident; if $c_0(R)$ is a solution then
$c_0( \lambda R)$ is also a solution for {\it any} value of $\lambda
\neq 0$. The solution must necessarily specify the value at a given
scale $c_0 (R_0)$, hence breaking explicitly the dilatation symmetry.
This symmetry breaking is unavoidable. In
Refs.~\cite{PavonValderrama:2004nb,PavonValderrama:2007nu} it is shown
how, for $g < -1/4 $ the breaking is lowered to the discrete subgroup
of dilatations, and the connection to the Russian Doll
renormalization.  In the case of the Wigner symmetry for the $^1S_0$
and $^3S_0$ potentials discussed in the paper, the breaking is not
unavoidable, and there exists in fact a very special choice where the
symmetry can be preserved by taking identical boundary conditions at a
given scale. Besides this particular solution, the identity between
solutions $c_{0,s}(R)$ and $c_{0,t}(R)$ will generally be violated,
although the relation from one scale to a different one $c_{0,s}(R_0)
\to c_{0,s}(R)$ and $c_{0,t}(R_0) \to c_{0,t}(R)$ is governed by the
{\it same} relation, Eq.~(\ref{eq:RG}).

It is worth noting the resemblance of the previous quantum-mechanical
discussion with similar and well-known field theoretical concepts.
The unavoidable breaking of the dilatation symmetry corresponds to an
anomaly of the dilatation current. The optional choice of boundary
conditions corresponds to the case of finite but ambiguous theories
(see e.g. Ref.~\cite{Jackiw:1999qq}).

\def\u{{\bf u}} \def\U{{\bf U}} \def\S{{\bf S}} \def\h{{\bf h}}
\def\L{{\bf L}} \def\E{{\bf 1}} \def\j{{\bf j}} \def\y{{\bf y}}
\def\M{{\bf M}} 
\def\f{{\bf f}} 

\section{Splitting formula for phase-shifts}
\label{sec:splitting}

We want to derive the splitting formula for phase shifts,
Eq.~(\ref{eq:ps-lande}) by using distorted waves perturbation theory.
The coupled channel Schr\"odinger equation for the relative motion
reads
\begin{eqnarray}
-\u '' (r) + \left[ \U (r) + \frac{{\bf L}^2}{r^2} \right] \u (r) =
 k^2 \u (r) \, , 
\label{eq:sch_cp} 
\end{eqnarray} 
where $\U^{SJ}_{L,L'} (r)= 2 \mu_{np} {\bf V}^{SJ}_{L,L'}(r)$ is the
coupled channel matrix potential which for the total angular momentum
$J> 0$ can be written as,
\begin{eqnarray}
\U^{0J} (r) &=& U_{JJ}^{0J} \nonumber \\ \\   
\U^{1J} (r) &=& \left( \begin{matrix} U_{J-1,J-1}^{1J} (r) & 0 &
U_{J-1,J+1}^{1J} (r) \cr 0 & U_{JJ}^{1J} (r) & 0 \cr U_{J-1,J+1}^{1J}
(r) & 0 & U_{J+1,J+1}^{1J} (r) \end{matrix} \right) \, \nonumber  
\end{eqnarray} 
In Eq.~(\ref{eq:sch_cp}) $ {\bf L}^2 = {\rm diag} ( L_1 (L_1+1),
\dots, L_N (L_N +1) )$ is the angular momentum, $\u(r)$ is the reduced
matrix wave function and $k$ the C.M. momentum. In the case at hand
$N=1$ for the spin singlet channel with $L=J$ and $ N=3 $ for the spin
triplet channel with $L_1=J-1$, $L_2=J$ and $L_3=J+1$. For ease of
notation we will keep the compact matrix notation of
Eq.~(\ref{eq:sch_cp}). At long distances, we assume the asymptotic
normalization condition
\begin{eqnarray}
\u (r)  \to \hat \h^{(-)} (r) - \hat \h^{(+)} (r) \S \, , 
\label{eq:asym}
\end{eqnarray} 
with $\S$ the standard coupled channel unitary S-matrix.  For the spin
singlet state, $S=0$, one has $L=J$ and hence the state is un-coupled
\begin{eqnarray}
S_{JJ}^{0J} = e^{ 2 i \delta_{J}^{0J} } \, , 
\end{eqnarray}
whereas for the spin triplet state $S=1$, one has the un-coupled $ L=J$
state
\begin{eqnarray}
S_{JJ}^{1J} &=& e^{  2 i \delta_{J}^{1J} } \;,
\end{eqnarray}
and the two channel coupled states $L,L'=j \pm 1$ states which written
in terms of the eigenphases are 
\begin{eqnarray}
S^{1J} &=& \left( \begin{matrix} \cos \epsilon_J & -\sin \epsilon_J \cr \sin
\epsilon_J & \cos \epsilon_J \end{matrix} \right) \left( \begin{matrix} e^{2 {\rm i}
\delta^{1J}_{J-1}} & 0 \cr 0 & e^{2 {\rm i} \delta_{J+1}^{1J}} \end{matrix}
\right) \nonumber \\ &\times& \left( \begin{matrix} \cos \epsilon_J & \sin
\epsilon_J \cr -\sin \epsilon_J & \cos \epsilon_J \end{matrix} \right) \, .
\label{eq:BB} 
\end{eqnarray} 
The corresponding out-going and in-going free spherical waves are
given by
\begin{eqnarray}
\hat \h^{(\pm)} (r) &=& {\rm diag} ( \hat h^\pm_{L_1} ( k r) , \dots ,
\hat h^\pm_{L_N} (k r) ) \, ,
\end{eqnarray} 
with $ \hat h^{\pm}_L ( x) $ the reduced Hankel functions of order
$l$, $ \hat h_L^{\pm} (x) = x H_{L+1/2}^{\pm} (x) $ ( $ \hat h_0^{\pm}
= e^{ \pm i x}$ ), and satisfy the free Schr\"odinger's equation for a
free particle. 

In order to determine the infinitesimal change of the $S$ matrix, $\S
\to \S + \Delta \S$, under a general deformation of the potential $\U
(r) \to \U(r) + \Delta \U(r) $ we use Schr\"odinger's equation
(\ref{eq:sch_cp}) and the standard Lagrange's identity adapted to this
particular case, we get
\begin{eqnarray}
\left[ \u (r)^\dagger \Delta \u'(r) - \u'(r)^\dagger \Delta \u(r)
\right]' = 
\u(r)^\dagger \Delta \U(r) \u(r) \, . \nonumber \\ 
\end{eqnarray} 
The unitarity of the S-matrix, $\S^\dagger \S= {\bf 1}$, yields the
condition $\Delta \S^\dagger \S+ \S^\dagger \Delta \S =0$.  We assume
a mixed boundary condition at short distances, $r=r_c$, for the
unperturbed coupled channel potential, $\U(r)$,
\begin{eqnarray}
\u' (r_c) + \L \u (r_c) = 0 \, ,  
\label{eq:bc}
\end{eqnarray}   
with $\L$ a self-adjoint matrix. After integration from the cut-off
radius $r_c$ to infinity and using the asymptotic form of the matrix
wave function, Eq.~(\ref{eq:asym}), as well as the condition at the
origin, Eq.~(\ref{eq:bc}) yields
\begin{eqnarray}
2 {\rm i} k \S^\dagger \Delta \S = \int_{r_c}^\infty dr \, \u(r)^\dagger
\Delta \U(r) \u(r) \, .
\end{eqnarray}
If we take the Wigner symmetric states as the unperturbed problem,
then $\S$ , $\U(r)$ and $\u(r)$ become a diagonal matrices, so that 
\begin{eqnarray}
\Delta \delta_{JL}^{ST} = - \frac1{2p} \int_{r_c}^\infty dr \,
u_L^{ST} (r)^\dagger \Delta \U(r) u_L^{ST} (r) \, , 
\end{eqnarray}
so that the perturbed eigenphases become
\begin{eqnarray}
\delta_{JL}^{ST} = \delta_{L}^{ST} + \Delta \delta_{JL}^{ST} 
\end{eqnarray}
Note that to this order the mixing phases vanish, $\Delta
\epsilon_J=0$. Identifying further $\Delta \U$ with the spin-orbit and 
the tensor potential, in the
LS-coupling the result may be written as
\begin{eqnarray}
\delta_{LJ}^{ST} &=& \delta_{L}^{ST} + \delta_{S,1} C_L^{ST} (S_{12}^{J})_{LL}   \nonumber \\ 
&+& A_L^{ST}  \left[ J(J+1)-L(L+1)-S(S+1) \right] 
\, , \nonumber \\
\end{eqnarray}  
where $ (S_{12}^{J})_{J-1,J-1} = -2(J-1)/(2J+1)$, $ (S_{12}^{J})_{J,J}
= 2 $, $ (S_{12}^{J})_{J+1,J+1} = -2(J+2)/(2J+1)$.  Defining the
supermultiplet coefficients $A_L= A_L^{10}= A_L^{01}$ and $B_L=
A_L^{00}= A_L^{11}$
\begin{eqnarray}
\delta_{LJ}^{10} &=& \delta_{LJ}^{01} + A_L \left[ J(J+1)-L(L+1)-2 \right]
\nonumber \\ &+& C_L (S_{12}^{J})_{LL} \, , \\ \delta_{LJ}^{11} &=&
\delta_{LJ}^{00} + B_L \left[ J(J+1)-L(L+1)-2 \right] \nonumber \\ &+& D_L
(S_{12}^{J})_{LL} \, , 
\end{eqnarray}  
we readily get the sum rule for phase-shifts,
Eq.~(\ref{eq:ps-lande}). The above equations would yield a Lande-like 
interval rule between spin-triplet energy levels for
the spin-orbit or the tensor potentials separately.  For instance,
\begin{eqnarray}
\delta_{^1P_1} &=& \delta_P \, , \nonumber  \\ \delta_{^3P_0} &=& \delta_P - 4 D_1 - 4
B_1 \, , \nonumber \\ \delta_{^3P_1} &=& \delta_P + 2 D_1 - 2 B_1 \, , \nonumber \\ \delta_{^3P_2}
&=& \delta_P - \frac{2}{5} D_1 + 2 B_1 \, .
\end{eqnarray} 
A further remark is in order, since the
spin-orbit or tensor potentials may be singular at the origin.  In
such a case of singular perturbations one computes the sum rule first
and then removes the cut-off, $r_c \to 0$.

\end{document}